\documentclass[ %
    reprint, superscriptaddress,
    amsmath,amssymb,
    aps,
    pra,
    ]{revtex4-2}
\usepackage{graphicx}
\usepackage{dcolumn}
\usepackage{bm}
\usepackage{orcidlink}

\usepackage{subcaption}
\usepackage{color}
\usepackage{comment}

\usepackage{graphics}
\usepackage{epsfig}
\usepackage{orcidlink}

\begin{document}
\preprint{APS/123-QED}

\title[Is the cortical dynamics ergodic?]{Is the cortical dynamics ergodic? \\ A numerical study in partially symmetric networks of spiking neurons}

\author{Ferdinand Tixidre\,\orcidlink{0009-0000-5065-7551}}
 \email{ferdinand.tixidre@cyu.fr}
 \affiliation{%
 Laboratoire de Physique Théorique et Modélisation, UMR 8089, CY Cergy Paris Université, CNRS, Cergy-Pontoise, France}%
 
\author{Gianluigi Mongillo}
\affiliation{Sorbonne Université, INSERM, CNRS, Institut de la Vision, F-75012 Paris, France.}%
\affiliation{Centre National de la Recherche Scientifique, Paris, France.}%
 
\author{Alessandro Torcini}%
\affiliation{%
 Laboratoire de Physique Théorique et Modélisation, UMR 8089, CY Cergy Paris Université, CNRS, Cergy-Pontoise, France}%
\affiliation{CNR - Consiglio Nazionale delle Ricerche - Istituto dei Sistemi Complessi, via Madonna del Piano 10, 50019 Sesto Fiorentino, Italy}
\affiliation{INFN Sezione di Firenze, Via Sansone 1, 50019 Sesto Fiorentino, Italy}

\date{\today}

\begin{abstract}
Cortical activity in-vivo displays relaxational time scales much longer than the membrane time constant of the neurons or the deactivation time of ionotropic synaptic conductances. The mechanisms responsible for such slow dynamics are not understood. Here, we show that slow dynamics naturally and robustly emerges in dynamically-balanced networks of spiking neurons. This requires only partial symmetry in the synaptic connectivity, a feature of local cortical networks observed in experiments. The symmetry generates an effective, excitatory self-coupling of the neurons that leads to long-lived fluctuations in the network activity, without destroying the dynamical balance. When the excitatory self-coupling is suitably strong, the same mechanism leads to multiple equilibrium states of the network dynamics. Our results reveal a novel dynamical regime of the collective activity in spiking networks, where the memory of the initial state persists for very long times and ergodicity is broken.
\end{abstract}
\maketitle

\section{Introduction}
\label{sec:intro}

In the active cortex, neurons exhibit significant correlations in their spike counts, even when these counts are separated by lags as long as 1 second or more \cite{ogawa2010differential,murray2014hierarchy,nishida2014discharge}. The dynamics of the membrane depolarization and the activation/deactivation of synaptic conductances have, instead, typical time scales on the order of $10$ms \cite{dayan2001theoretical}. Hence, these processes cannot account for the long-lived fluctuations of spiking activity observed in experiments.

Theoretically, there are two "minimal" scenarios to account for them. In one scenario, the long time scales are simply inherited from some slow underlying biophysical processes. For instance, both neurons \cite{la2006multiple,pozzorini2013temporal} and synapses \cite{fisher1997multiple} exhibit dynamics on multiple time scales, some of the order of $1$-$10$s. We are not aware of any systematic investigation of this scenario. We note, however, that the relationship between the time scales of the microscopic (neuron/synapse) and the macroscopic (network) dynamics is far from trivial; slow microscopic dynamics does not automatically result in slow macroscopic dynamics \cite{beiran2019}. In the other scenario, the long time scales emerge from the interaction of fast biophysical processes. In the following, we focus on this scenario.

The theory of balanced networks \cite{tsodyks1995rapid,van1996,amit1997model,renart2010asynchronous} elegantly accounts for many non-trivial features of the cortical activity. According to this theory, in an active cortical network, the excitatory and the inhibitory drive to single neurons are both large, but the level of activity in the network dynamically adjusts so that the net drive is around threshold. This leads to a dynamical state---the balanced regime---characterized by temporally-irregular spiking at the single-neuron level with low, heterogeneous firing rates  \cite{lerchner2006response,roxin2011distribution,mongillo2012bistability,mongillo2018inhibitory} and by weak pairwise correlations in the spiking of different neurons \cite{renart2010asynchronous,helias2014correlation}, reproducing many characteristics of the patterns of activity observed in the cortex. Long relaxation times, however, are not a generic feature of {\em random} networks operating in the balanced regime; additional mechanisms have to be postulated. 

There are presently three main proposals to explain the emergence of long time scales in cortical networks. According to one proposal, cortical networks operate close to the {\em edge of stability} \cite{harish2015asynchronous,kadmon2015,dahmen2019second}. Depending on the level of variability in the synaptic efficacies (synaptic gain), random networks of {\em rate} neurons display either a unique, stable fixed point or a chaotic attractor. Adjusting the synaptic gain suitably close to its critical level, one can achieve arbitrarily slow relaxational dynamics. As the network keeps operating in the balanced regime on both sides of the transition, pairwise correlations remain weak.

However, such a transition {\em does not} exist in spiking networks \cite{harish2015asynchronous,kadmon2015}. More precisely, a "sharp" transition is recovered only when the {\em synaptic} time constant becomes infinitely long, so that spiking neurons effectively operate as rate elements \cite{harish2015asynchronous,kadmon2015,angulo2017}. Considering typical firing rates ($\sim 1$Hz) and synaptic time constants ($10-100$ms) in the cortex, one expects a significant "smoothing" of the transition; in particular, the autocorrelation time remains finite regardless of the level of the synaptic gain \cite{kadmon2015}. In fact, numerical simulations show that the autocorrelation time remains of the order of the synaptic time constant when the network parameters, in particular synaptic time constants, are chosen within ranges of biological credibility \cite{harish2015asynchronous}.    

According to another proposal, cortical networks are {\em metastable} \cite{deco2012neural,litwin2012,mazzucato2015dynamics}; see Ref. \cite{brinkman2022metastable} for a general overview. With suitable synaptic structuring, model networks feature multiple steady states of activity, differing in the spatial distributions of "active" (high-rate) and "inactive" (low-rate) neurons. {\em Noise}, then, induces transitions between these states. Thus, single neurons exhibit fluctuations in their firing rate on time scales comparable with the lifetime of network's steady states. For suitably low levels of noise, these lifetimes can be significantly longer than the microscopic time constants. Furthermore, if pairs of neurons are co-active only in a small fraction of the steady states (sampled by the dynamics during the typical duration of the experiments), measured pairwise correlations remain weak \cite{litwin2012,mazzucato2015dynamics}.

Achieving levels of noise suitable to match the experimental observations, however, requires significant fine-tuning; see Ref. \cite{huang2017once} for a discussion of this point. The reason is easy to understand. Life-times much longer than the microscopic time scales require the existence of quasi-equilibrium states {\em\`a la} Kramers \cite{hanggi1990}. Accordingly, the transition rate between the different states will, to a good approximation, depend exponentially on the level of noise; see, e.g., Ref \cite{litwin2012}. The level of noise, in turn, is controlled by the number of "active" neurons in a steady state, the strength of the recurrent synaptic couplings and the level of the external drive to the network. It is presently unclear, if neural and/or synaptic plasticity mechanisms can control these parameters with the required, exponential, precision; see, e.g., Ref \cite{yang2024co} for a proposal based on synaptic plasticity.   

A more recent and, hence, much less investigated proposal is that long time scales result from the symmetry in the synaptic connectivity \cite{marti2018,rao2019,berlemont2022glassy}. In fact, local connectivity (i.e., on spatial scales of the order of 100 $\mu$m) in the cortex is not random. Rather, reciprocal synaptic connections are both over-represented, as compared to random networks, and stronger than average \cite{song2005,yoshimura2005,ko2011,brunel2016cortical,campagnola2022local}. On first principles, then, one would expect a slowing down of the dynamics because of the {\em effective} positive self-coupling (i.e., the so-called {\em reaction term}; see Sec.~\ref{sec:results}) generated by the symmetry. Indeed, the slowing down is observed in model networks, both of rate \cite{crisanti1988dynamics,marti2018,berlemont2022glassy} and of spiking neurons \cite{rao2019}. A detailed investigation of the effects of symmetry on the dynamics of spiking networks, however, is still lacking. We provide here such an investigation.

We consider networks of inhibitory leaky-integrate-and-fire neurons, with arbitrary levels of symmetry in the synaptic connectivity and operating in the balanced regime, as a {\em minimal} model of a local cortical network. At increasing levels of symmetry, and for suitably large synaptic efficacies, we find that the neurons display large and long-lived fluctuations in their spiking rate, similarly to what is observed in {\em clustered} model networks and in experiments \cite{deco2012neural,litwin2012,mazzucato2015dynamics}. As the symmetry level is increased, the lifetime of the fluctuations becomes comparable to the duration of our simulations ($\sim 10^{4}$ seconds). Notably, this behavior is independent of the synaptic time constant and occurs in partially symmetric networks. In the parameter region where these exceptionally long-lived fluctuations exist, {\em interpolating} between spiking and rate neurons by progressively increasing the synaptic time constant, we find that the network dynamics exhibit multiple, {\em stable} fixed points. 

Taken together, our results suggest that, for suitable levels of synaptic symmetry and of synaptic efficacies, the dynamics of a balanced network of spiking neurons is non-ergodic. In this parameter region, different initial conditions lead to different asymptotic states of activity.

\begin{figure*}[hbt!]
    \includegraphics[width=\textwidth]{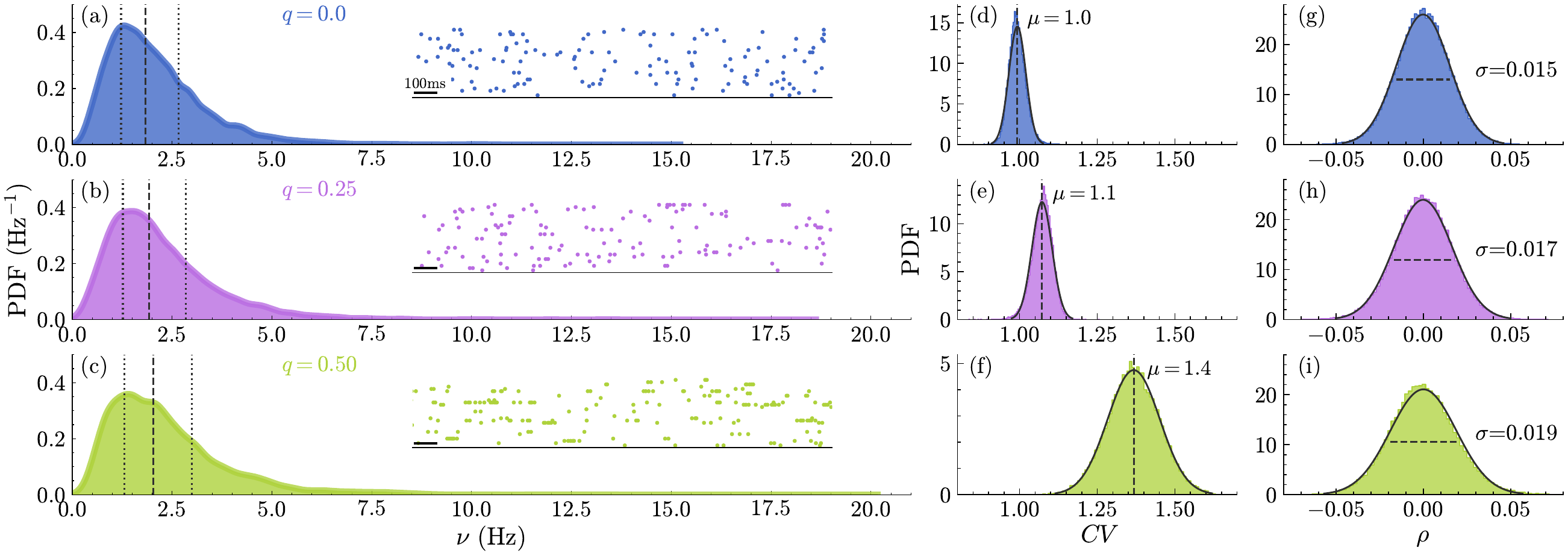}
    \caption{(a)–(c) PDFs of the single-neuron firing rates $\{\nu_i\}$. Dashed lines denote the mean and the dotted lines represent the first and third quartiles. The insets display the activity of thirty randomly selected neurons. (d)–(f) PDFs of the coefficients of variation $\{CV_i\}$ of individual neuron firing rates. The dotted vertical  lines indicate the means of the PDFs. (g)–(i) PDFs of the pairwise correlation coefficients $\rho_{ij}$ of the neuronal firing activity. The dotted horizontal lines indicate the full widths at half maximum of the PDFs.  The solid black curves in (d)–(f) and (g)–(i) represents Gaussian fits to the data, the corresponding mean and standard deviation are denoted by $\mu$ and $\sigma$, respectively. The top row (blue) corresponds to $q=0.0$; the middle row (magenta) to $q=0.25$ and the bottom one (green) to $q=0.5$. The PDFs were obtained by aggregating data  from 10 independent replicas. Simulation parameters: $N=4000$, $K=1200$, $g=5$, and $h_\mathrm{ext} = 0.10$.}
    \label{fig:Lif_delta_NetworkActivity} 
\end{figure*}

\section{Results}
\label{sec:results}

As a paradigmatic example of a network operating in the balanced regime, we consider $N$ randomly and sparsely connected inhibitory neurons receiving a constant, excitatory input from the outside \cite{monteforte2010,kadmon2015,harish2015asynchronous,di2018}. Single-neuron dynamics is described by the standard leaky integrate-and-fire (LIF) mechanism \cite{dayan2001theoretical}. Each neuron randomly makes $K\left(\ll N\right)$ synaptic contacts, on average, with the other neurons. In order to obtain a dynamically-balanced dynamics \cite{van1996,renart2010asynchronous}, we assume that the efficacy of the synaptic contacts is $-g/\sqrt{K}$ with $g>0$, so that the recurrent synaptic input is proportional to $\sqrt{K}$ and negative, and that the external excitatory input to each neuron is $h_\mathrm{ext}\sqrt{K}$ with $h_\mathrm{ext}>0$.

We control the level of symmetry in the network by manipulating the probability of reciprocal connection between pairs of neurons \cite{rao2019}. The {\em excess reciprocity}, $\eta$, defined as the ratio between the probability of finding reciprocally connected neurons and the probability of reciprocal connections in a random network (i.e., directed Erd\H{o}s-R\'enyi with the same probability of connection $K/N$) is given by (see Sec.~\ref{sec:methods} for details)

\begin{equation}
\eta=\frac{q}{K/N}+(1-q)^2    
\end{equation}

\noindent where $0\leq q\leq 1$. For $q=0$, the network is simply a random network ($\eta=1$), while for $q=1$ the network is fully symmetric ($\eta=N/K$; all connections are reciprocal). Experimental estimates of the probability of local connections ($0.1$-$0.3$) and of the excess reciprocity ($3$-$5$) in the cortex \cite{campagnola2022local} lead to $0.4\leq q\leq 0.9$. In particular, we note that the inhibitory-to-inhibitory network in the cortex is close to fully symmetric, i.e., $0.8\leq q\leq 0.9$.      

We start by considering the case of instantaneous synaptic currents (see Sec.\ref{sec:model} for details), i.e., when a neuron fires a spike, it instantaneously {\em reduces} the voltage of its postsynaptic targets by $g/\sqrt{K}$. In the absence of pairwise correlation in the synaptic connectivity (i.e., $q=0$), our model network operates in the balanced regime when $K\gg1$. In this regime, the large excitatory drive (the external input) is dynamically offset by the comparably large inhibitory drive (the recurrent input), resulting in the unsaturated, temporally irregular, and asynchronous spiking of the neurons in the network \cite{van1996,renart2010asynchronous}. This regime is illustrated in Fig.~\ref{fig:Lif_delta_NetworkActivity}. In Fig.~\ref{fig:Lif_delta_NetworkActivity}(a) we show the distribution of firing rates in the network; as can be seen it is right-skewed and long-tailed, and, it can be well approximated by a log-normal distribution (not shown) \cite{roxin2011distribution,mongillo2018inhibitory}. Moreover, the population-averaged firing rate, $\langle \nu \rangle = 2.1$Hz, is in good agreement with the value, $\nu_{\rm B}$, predicted by the balance condition in the limit $K\to\infty$, that is, $\nu_{\rm B}={h_\mathrm{ext}}/(\tau_m g) = 2$ Hz where $\tau_m=10$ ms is the membrane time constant \cite{monteforte2010,di2018}. As expected in the balanced regime, spiking is irregular, as assessed by the coefficient of variation (CV) of the inter-spike intervals of single neurons (Fig.~\ref{fig:Lif_delta_NetworkActivity}(d)), and asynchronous, as assessed by the zero-lag cross-correlation of the spike counts of different neurons (Fig.~\ref{fig:Lif_delta_NetworkActivity}(g)) \cite{renart2010asynchronous,helias2014correlation}.  

Increasing the levels of symmetry up to $q=0.5$ does not affect these features qualitatively; spiking activity remains unsaturated, temporally irregular and asynchronous. There are some quantitative changes, however. First, the distribution of firing rates becomes more skewed while the population-averaged firing rate remains approximately constant, as can be seen by comparing Fig.~\ref{fig:Lif_delta_NetworkActivity}(b-c) with Fig.~\ref{fig:Lif_delta_NetworkActivity}(a): $\langle \nu \rangle$ varies from $2.1$ Hz for $q=0$ to $2.3$ Hz for $q=0.5$. Second, the temporal irregularity of the spike trains increases, as can be seen in Fig.~\ref{fig:Lif_delta_NetworkActivity}(e-f); the population-averaged coefficient of variation, $\langle CV \rangle$, increases from $1$ for $q=0$ to $1.4$ for $q=0.5$. However, there are no noticeable changes in the level of "asynchrony", as can be seen in Fig.~\ref{fig:Lif_delta_NetworkActivity}(h-i). Thus, the network continues to operate in the balanced regime.

\begin{figure*}[hbt!]
    \centering
    \includegraphics[width=\linewidth]{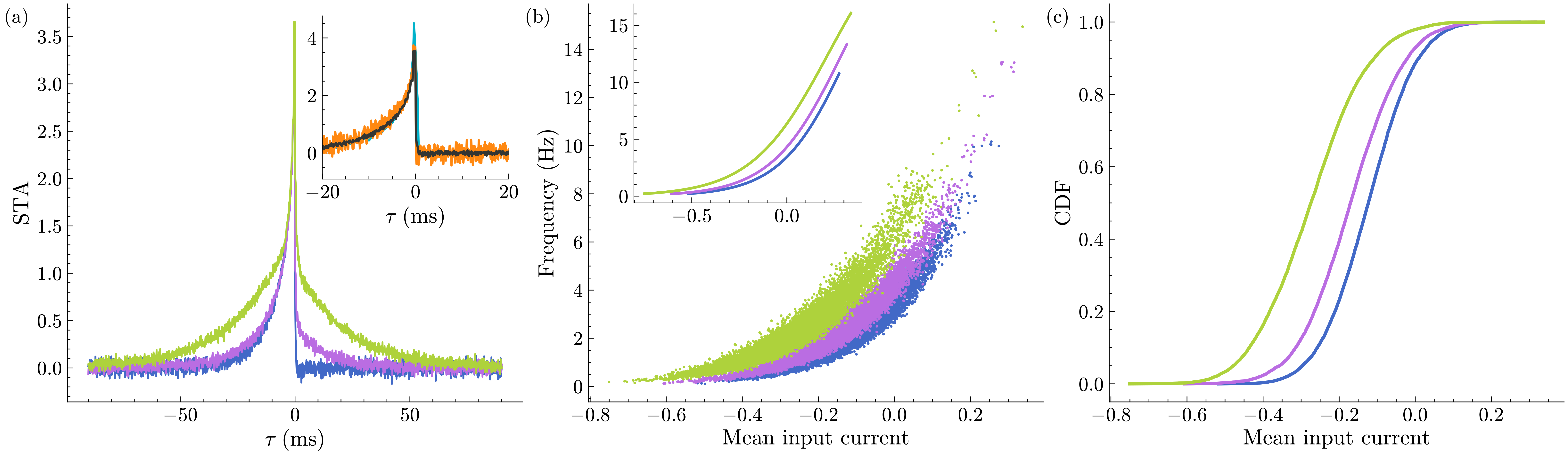}
    \caption{(a) Spike-triggered average input ${\rm STA}(\tau)$ for $q=0.0$ (blue), $q=0.25$ (magenta) and $q=0.5$ (green). In the inset of (a) we compare the STA obtained from the network simulations (grey), with a single neuron's dynamics under the diffusive approximation \cite{capocelli} (sky blue) and subject to Poissonian shot noise (orange). (b) Firing rates of the neurons versus the time averaged (mean) input currents for different levels of symmetry in the network $q=0.0$ (blue), $q=0.25$ (magenta), $q=0.5$ (green); the Population Frequency-Input curves are shown in the inset and obtained by fitting the logistic function to the raw data (details in \cite{sigmoid_fit}). (c) Cumulative distribution functions (CDFs) of the mean input currents.  The curves reported in (a) were obtained by averaging over $200,000$ spike events, while data in (b) and (c) has been obtained  by averaging over a time interval of 250 seconds. Same parameters used as in Fig.\ref{fig:Lif_delta_NetworkActivity}}
    \label{fig:ResponseToInput}
\end{figure*}

As we shall explain shortly, these quantitative changes are a consequence of the {\em effective} positive feedback generated by the symmetry in the synaptic connectivity (see also Ref.~\cite{rao2019}). First, let us discuss the mechanism generating the positive feedback. 

Consider neuron $i$; every time it fires a spike, it delays spiking in its postsynaptic targets. Indeed, upon spiking, neuron $i$ reduces the voltage of its postsynaptic targets by $g/\sqrt{K}$ and, hence, transiently reduces their instantaneous probability of spiking by a factor $\sim 1/\sqrt{K}$ (for $K \gg 1$). This is simply a consequence of the fact that the neurons are inhibitory. In the presence of symmetry, many of the postsynaptic targets of neuron $i$ are also its presynaptic neurons, that is, a fraction $q$ (on average). A reduction $\sim 1/\sqrt{K}$ in the probability of spiking of one of these neurons, then, leads to a reduction $\sim g/K$ in the (average) inhibitory input to neuron $i$. As there are $\sim qK$ of them, the total reduction is $\sim 1$ (i.e., independent of $K$, when $K\gg 1$). As a result, the {\em net} synaptic input received by neuron $i$ {\em following} its spike is transiently increased, because of the (transiently) reduced probability of spiking of its inhibitory presynaptic neurons. The typical size of this positive feedback is $\sim 1$ and, hence, the balanced regime is unaffected by it \cite{rao2019}.

The presence of such a positive feedback is readily revealed by the analysis of the spike-triggered average input, ${\rm STA}(\tau)$ (see Sec.~\ref{sec:indicators} for details), where $\tau$ is time relative to spike emission ($\tau=0$). Let us consider first the case $q=0$, where no positive feedback is expected. The corresponding ${\rm STA}(\tau)$ is the blue curve in Fig.~\ref{fig:ResponseToInput}(a). As can be seen, ${\rm STA}(\tau)$ increases above baseline briefly {\em before} the spike (i.e., $\tau < 0$), reflecting the fact that spikes are generated by suitably large, positive fluctuations in the input, and drops back to baseline immediately {\em after} the spike (i.e., $\tau>0$), indicating that there is no {\em consistent} effect of the spike on subsequent inputs. As can be seen in the inset of Fig.~\ref{fig:ResponseToInput}(a), this is the same ${\rm STA}(\tau)$, for all practical purposes, that one obtains from a LIF neuron driven by white noise \cite{capocelli} or by shot noise (i.e., a Poissonian spike train inducing finite-amplitude postsynaptic potentials), where, by construction, there is no correlation between spikes and synaptic inputs. 

\begin{figure}[hbt!]
\centering
\includegraphics[width=\linewidth]{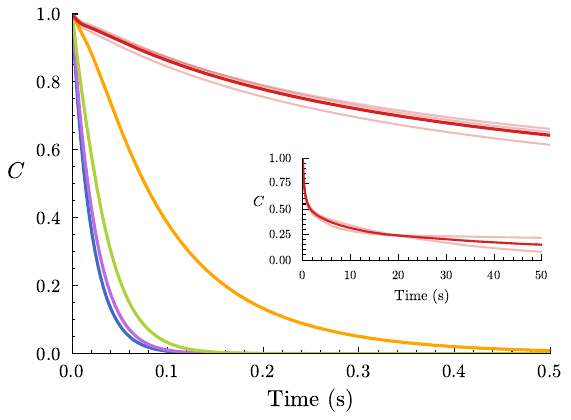}
\caption{The autocorrelations are shown as a function of the time lag for different levels $q$ of symmetry in the synaptic connections: $q=0.0$ (blue), $q=0.25$ (magenta), $q=0.5$ (green), $q=0.8$ (orange) and $q=0.9$ (red). The inset shows a zoomed-in view for $q=0.9$, where different shades of the red color denotes different initial conditions. The autocorrelations were averaged over 500 selected neurons and over 3 replicas. The same parameters as in Fig.\ref{fig:Lif_delta_NetworkActivity} were used.}
\label{fig:Panel3_Correlations}
\end{figure}

As expected, the situation changes when the level of symmetry is increased. For $q=0.25$, we see that ${\rm STA}(\tau)$ remains significantly above the baseline for about $20$ ms after the spike (magenta curve in Fig.~\ref{fig:ResponseToInput}(a)). This "after-spike transient" in the synaptic input is further amplified, both in amplitude and duration, when $q=0.5$ (green curve in Fig.~\ref{fig:ResponseToInput}(a)). It is convenient to quantify the strength of the positive feedback, $R$, by integrating the spike-triggered  average input for positive $\tau$ \cite{dim}, i.e., 
\begin{equation}
R\equiv\intop_{0}^{+\infty} \mathrm{d}\tau \;\mathrm{STA}\left(\tau\right)
\end{equation}
For $q=0$, we find $R \simeq 0.1$ ms, while $R \simeq 6$ ms for $q=0.25$ and $R \simeq 24$ ms for $q=0.5$.

With Fig.~\ref{fig:ResponseToInput}(a) in mind, it becomes easier to understand the effects of the symmetry on the patterns of spiking activity, illustrated in Fig.~\ref{fig:Lif_delta_NetworkActivity}. Let us start with the temporal structure of the spike trains. The presence of the positive feedback makes spiking a self-exciting process: firing a spike increases, transiently, the probability of firing another spike shortly. The strength of the self-excitation (i.e., $R$ defined above) increases with increasing levels of symmetry (see Fig.~\ref{fig:ResponseToInput}(a)) which, in turn, leads to an increasing tendency to fire spikes in "bursts". As a result, the mean coefficient of variation $\langle CV \rangle$ increases with $q$.

Next, let us consider the effect on the distribution of firing rates. At parity of synaptic input, the firing rate of a neuron increases with the level of symmetry $q$ (and hence $R$). This is illustrated in Fig.~\ref{fig:ResponseToInput}(b), where we plot the firing rates of the neurons in the network as a function of their time-averaged synaptic input, for the same values of $q$ (and using the same color coding) as in Fig.~\ref{fig:Lif_delta_NetworkActivity}. In the inset, we plot the population Frequency-Input (F-I) curves, obtained by fitting the data in the main panel with a logistic function \cite{sigmoid_fit}. 
 
\begin{figure*}[hbt!]
    \centering
    \includegraphics[width=\linewidth]{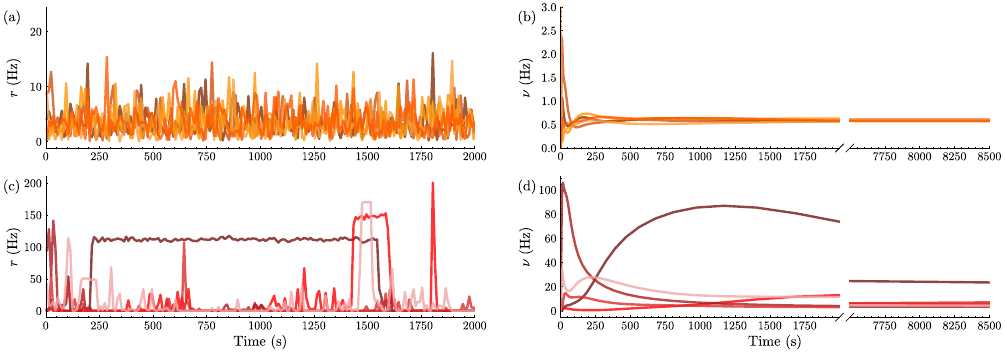}
    \caption{(a),(c) The "instantaneous" firing rate for the same neuron and replicas, using time bins of $\Delta  = 10$ seconds, for $5$ different replicas for $q=0.8$ (orange) and $q=0.9$ (red). Different shades of the main color represent different replicas. (b),(d) Running averages of the firing rate of a single selected neuron versus time. All other parameters are the same as in Fig. \ref{fig:Lif_delta_NetworkActivity}.}
    \label{fig:RatesInReplicas}
\end{figure*}

However, the population-averaged firing rate $\langle \nu \rangle$ cannot change dramatically. This is due to the fact that $\langle \nu \rangle$ is essentially fixed (i.e., within $\mathcal{O}\left(1/\sqrt{K}\right)$) to $\nu_{\rm B}$ by the balance condition, and the value of $\nu_{\rm B}$ is independent of $q$. To "solve" this apparent contradiction, the distribution of the firing rates adjusts so as to reduce the time-averaged synaptic inputs while (approximately) preserving the population-averaged firing rate. This reduction is illustrated in Fig.~\ref{fig:ResponseToInput}(c). The reduction of the time-averaged synaptic inputs leads to an increase in the fraction of both low-firing and high-firing neurons with increasing $q$, as clearly visible in the inset of Fig.~\ref{fig:ResponseToInput}(b).

As a consequence of the increased tendency of neurons to fire spikes in bursts, we expect large, positive correlations between the synaptic inputs (to the same neuron) at time lags longer than the membrane time constant ($\tau_m = 10$ ms). We also expect these correlations to persist over longer lags as  $q$ increases. This is confirmed in Fig.~\ref{fig:Panel3_Correlations}, where we plot the population-averaged autocorrelation function of the synaptic inputs, $C\left(t\right)$, for different values of $q$ (see Sec.~\ref{sec:indicators} for details).

As can be seen, consistent with the expectations, $C(t)$ is non-negative, and the time it takes to decay to $0$ gets longer with increasing $q$. For levels of symmetry up to $q=0.8$, $C(t)$ is well fitted by a single exponential decay, i.e., $C(t)=e^{-t/\tau_C}$ for $t\geq 0$. The time constant $\tau_C$, extracted from the fit, provides a simple quantification of the decay time. For $q=0$ (blue curve), we find $\tau_C\simeq20$ ms, while for $q=0.8$ (orange curve), we find $\tau_C\simeq100$ ms. 

At $q=0.9$ (red curve), however, the behavior of $C\left(t\right)$ changes dramatically. As can be seen in the inset of Fig.~\ref{fig:Panel3_Correlations}, $C\left(t\right)$ remains significantly different from $0$ at $t=50$ s; three orders of magnitudes larger than $\tau_m$. Also, the decay of $C\left(t\right)$ is clearly different from a single exponential decay; indeed, it features at least two time scales. Finally, $C\left(t\right)$ displays noticeable fluctuations even at short lags when re-estimated (i.e., in the same network) starting from different initial conditions (lighter-red curves). For $q \leq 0.8$, the same re-estimation procedure produces virtually identical $C\left(t\right)$ (not visible in Fig.~\ref{fig:Panel3_Correlations} since they are essentially overlapping).    
\begin{figure*}[hbt!]
    \centering
    \includegraphics[width=\linewidth]{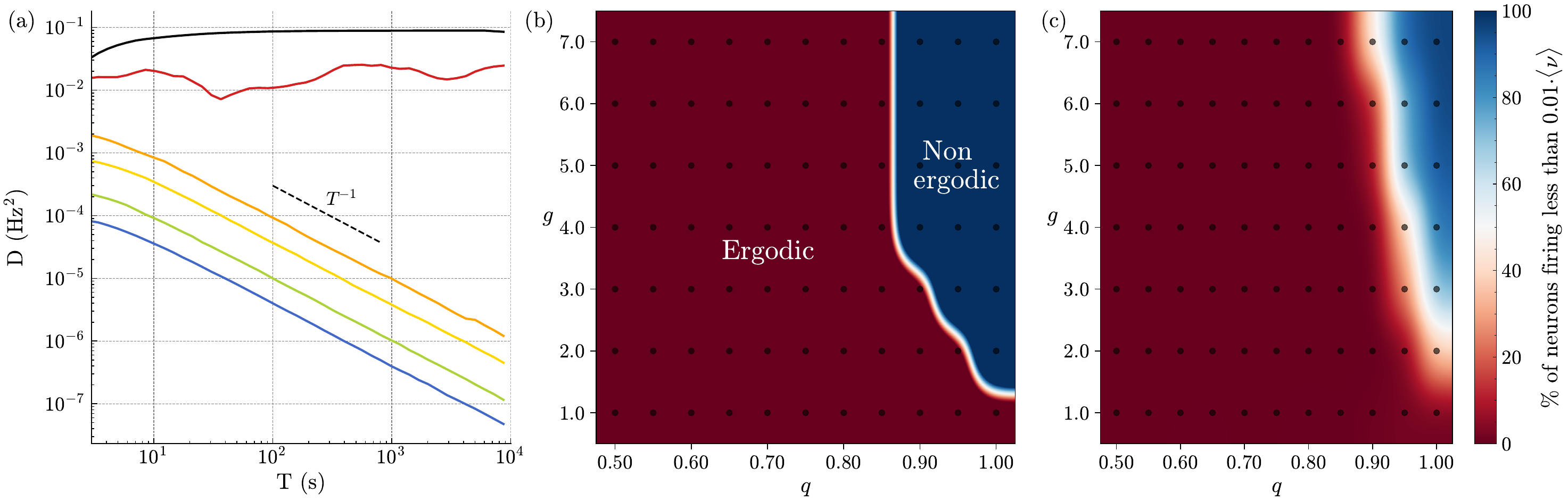}
    \caption{(a) $D$ as a function of time for different values of $q$: $q=0$ (blue), $q=0.5$ (green), $q=0.7$ (yellow), $q=0.8$ (orange),  $q=0.9$ (red), $q=1.0$ (black). Five realizations (replicas) of the same network have been considered to estimate $D$. (b) Bidimensional phase diagram in the $(q,g)$ plane, showing the ergodic and non-ergodic regimes in the red and blue regions, respectively.  (c) Percentage of neurons with a firing rate lower than $0.01 \langle \nu \rangle$ (i.e., almost silent).  The black dots in (b-c) indicate the parameter values for which the reported quantities in the plane have been measured.  All other parameters are fixed as in Fig. \ref{fig:Lif_delta_NetworkActivity}.}
    \label{fig:ErgodicityBreaking}
\end{figure*}

Thus, the analysis of $C\left(t\right)$ reveals a {\em qualitative} change in the network dynamics, occurring between $q=0.8$ and $q=0.9$. For $q=0.8$, the fluctuations in the activity decay exponentially at long time lags, suggesting that the network dynamics has reached a stationary state. For $q=0.9$, the fluctuations in the activity do not decay exponentially; rather, they persist for a time comparable with the duration of the simulations, suggesting that the network dynamics has not reached a stationary state. 

To better understand the dynamical mechanism responsible for such long memory in the fluctuations, we simulated the dynamics in the {\em same} network starting from {\em different} initial conditions, that is, different values of neurons' membrane potentials at time $t=0$. 

In Fig.~\ref{fig:RatesInReplicas}(a), we plot the time course of the firing rate of a randomly chosen neuron for $5$ different initial conditions, at $q = 0.8$. The firing rate is estimated in consecutive, non-overlapping bins of duration $10$s. The $5$ corresponding time series appear "stationary", consistent with the analysis of $C\left(t\right)$ (see Fig.~\ref{fig:Panel3_Correlations}). Indeed, this is expected if the spike trains for the different initial conditions were simply different realizations of the same "underlying", constant (i.e., on time scales longer than $\tau_C$) firing rate. To qualify this statement, for each initial condition, we have estimated the firing rate by counting all the spikes occurring between $t = 0$ and $t = T$. As can be seen in Fig.~\ref{fig:RatesInReplicas}(b), these estimates converge to the same value at increasing $T$, regardless of the initial condition.    

The same analysis for $q=0.9$ is shown in Fig.~\ref{fig:RatesInReplicas}(c). In this case, the time series for the different initial conditions clearly do not resemble different realizations of the same "underlying", constant firing rate. One can see large (up to $\sim 100$ Hz) and very long-lived (up $\sim 1000$ s) fluctuations in the firing rates (estimated in $10$ s bins). It is not even clear that one can meaningfully define the "firing rate" of this neuron. Indeed, the procedure of estimating the firing rate, as above, by counting spikes in increasingly long time windows returns different values depending on the initial condition (Fig.~\ref{fig:RatesInReplicas}(d)).  

By definition, the dynamics is ergodic when the long-time average of any (sufficiently well-behaved) observable asymptotically converges to the same value, independently of the initial condition. In our case, the observable is the number of spikes in a time window $T$, and its long-time average is the "firing rate", that is, the probability of observing a spike per unit of time. As just shown, this property is satisfied for $q=0.8$, and it is not satisfied for $q=0.9$. The foregoing analysis, thus, reveals that the network dynamics is {\em not} ergodic at $q=0.9$.

We use the asymptotic convergence of firing rates to systematically study ergodicity breaking as a function of the level of symmetry, $q$, and of the synaptic efficacy, $g$. Specifically, we define a function of $T$, $D\left(T\right)$, which quantifies the difference among the firing-rate estimates, over a time window $T$, obtained from different initial conditions (see Sec.~\ref{sec:indicators} for details). We classify the dynamics as ergodic when $D\left(T\right)$ decays as $1/T$ for $T\to\infty$ \cite{gardiner1985handbook}, and as non-ergodic otherwise. Note that changes in $g$ at fixed $h_\mathrm{ext}$, the strength of the external drive, result in changes in the population-averaged firing rate, $\langle\nu\rangle$. In the region of the $\left(q,g\right)$ plane that we have explored numerically, $\langle\nu\rangle$ ranges between about $1$ Hz (for $g=7$) and about $5$ Hz (for $g=2$).

Consistent with the results illustrated in Figs.~\ref{fig:Panel3_Correlations} and \ref{fig:RatesInReplicas}, $D\left(T\right)\to 0$ as $1/T$ for $q \leq 0.8$, while for $q \geq 0.9$, $D\left(T\right)$ does not decay, even when firing rates are estimated over time intervals as long as $T=10^4$ s (see Fig.~\ref{fig:ErgodicityBreaking}(a)). In fact, as illustrated in Fig.~\ref{fig:ErgodicityBreaking}(b), ergodicity breaking occurs in a whole region of the $\left(q,g\right)$ plane. As can be seen, ergodicity breaking requires large values of $q$, the level of symmetry in the synaptic connectivity. However, symmetry alone is not sufficient; synaptic efficacies have to be large enough. For instance, for $g < 1.25$, the network dynamics is ergodic at $q=1.0$ (Fig.~\ref{fig:ErgodicityBreaking}(b)).

The breaking of ergodicity is accompanied by the appearance of "silent" neurons. This is illustrated in Fig.~\ref{fig:ErgodicityBreaking}(c), where we report the fraction of neurons firing at less than $0.01 \langle \nu \rangle$ as a function of $q$ and $g$. These neurons would be hardly detectable in experiments \cite{shoham2006silent}. As can be seen, for $q \geq 0.9$ and $g \geq 3$, more than half of the neurons in the network are "almost silent" (up to 87\% at $q=1.0$). By contrast, there are no silent neurons at $q=0.7$. The identity of the almost silent neurons and hence, of the active ones changes depending on the initial condition. In other words, in the region where ergodicity is broken, the network dynamics exhibits multiple, (very) long-lived states of activity differing in their spatial distribution of "active" and "silent" neurons. The fact that we were able to make such an observation by using a small number of initial conditions (i.e., $5$) suggests that the network dynamics can, indeed, exhibit a very large number of such states. 

What is the mechanism responsible for ergodicity breaking in our model network? Generally speaking, ergodicity breaking results from the fact that the dynamics is for very long times, or even permanently, restricted to different regions of the phase space depending on the initial conditions. Perhaps the simplest scenario for the emergence of such "trapping regions" is the co-existence of multiple, stable equilibria, e.g., different fixed points of the microscopic dynamics. However, our network does not have fixed points {\em stricto sensu}, i.e., states in which the voltage of all neurons is constant in time; this would correspond to a completely silent network. To assess the relevance of the above scenario in our model network, therefore, we introduce a synaptic time constant, $\tau_s$ (see Sec.~\ref{sec:model} for details). Roughly speaking, this leads to synaptic inputs proportional to the "average", rather than the "instantaneous", presynaptic activity, effectively reducing their (fast) temporal fluctuations. The time window over which the presynaptic activity is averaged is controlled by $\tau_s$. The network we have studied so far corresponds to $\tau_s=0$. In the opposite limit, i.e., $\tau_s \to \infty$, the fast fluctuations due to spiking disappear, and the dynamics can be described by a rate model \cite{kadmon2015,harish2015asynchronous}. Note that the time-averaged input to a neuron does not depend on $\tau_s$. 

We start by investigating how the ergodic properties of the network dynamics are affected by $\tau_s$. To this end, we compute the ergodic distance, $D\left(T\right)$, for different values of $\tau_s$ and $q$, for $g=5$. For this value of $g$, ergodicity is broken between $q=0.8$ and $q=0.9$ in the network with $\tau_s=0$ (see Fig.~\ref{fig:ErgodicityBreaking}). To properly compare the behavior of $D\left(T\right)$ at different values of $\tau_s>0$, we measure time in units of $\tau_s$, i.e., $\tilde{T}=T/\tau_s$. 

\begin{figure}[hbt!]
\centering
\includegraphics{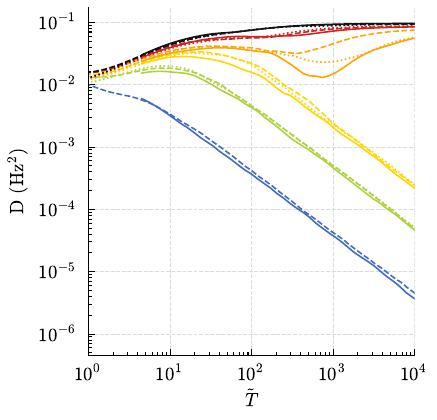}
\caption{$D$ as a function of $\tilde{T}$ for exponentially decaying synapses with different values of $q$ and of the synaptic time constant: namely, $\tau_s = 0.02$ s  (solid lines); 0.1 s (dashed lines); and 1 s (dotted lines). The colors  refer to $q=0$ (blue), $q=0.5$ (green), $q=0.7$ (yellow), $q=0.8$ (orange),  $q=0.9$ (red), and $q=1.0$ (black). These results were obtained using five replicas; other parameters are fixed as in Fig. \ref{fig:Lif_delta_NetworkActivity}}
\label{fig:synfilter}
\end{figure}

If the breaking of ergodicity in our model network is associated with the appearance of stable fixed points in the "corresponding" rate model (i.e., the same network in the limit $\tau_s\to\infty$), then one would expect that the region in the parameter space in which ergodicity is broken does not shrink with increasing $\tau_s$; rather, it could increase. This is indeed what we observe. As can be seen in Fig.~\ref{fig:synfilter}, the dynamics remain non-ergodic for $q\geq0.9$, regardless of the value of $\tau_s$. However, now the dynamics is non-ergodic already at $q=0.8$ for $\tau_s=0.02$s, and remains so by further increasing $\tau_s$. By contrast, the dynamics remain ergodic at $q=0.7$ even for $\tau_s=1$s. We conjecture that this is because there are no stable fixed points, in the limit $\tau_s\to\infty$, in the network at $q=0.7$. However, it is also conceivable that, instead, stable fixed points will appear for $\tau_s>1$s, and hence ergodicity will be broken for larger values of $\tau_s$. For practical reasons, i.e., simulations run times, we have systematically explored only $\tau_s\leq1$s. In any case, we do not expect the dynamics to become non-ergodic at $q=0$ because there are no {\em stable} fixed points in the network, even in the limit $\tau_s\to\infty$.     

\begin{figure*}[hbt!]
\includegraphics[width=\linewidth]{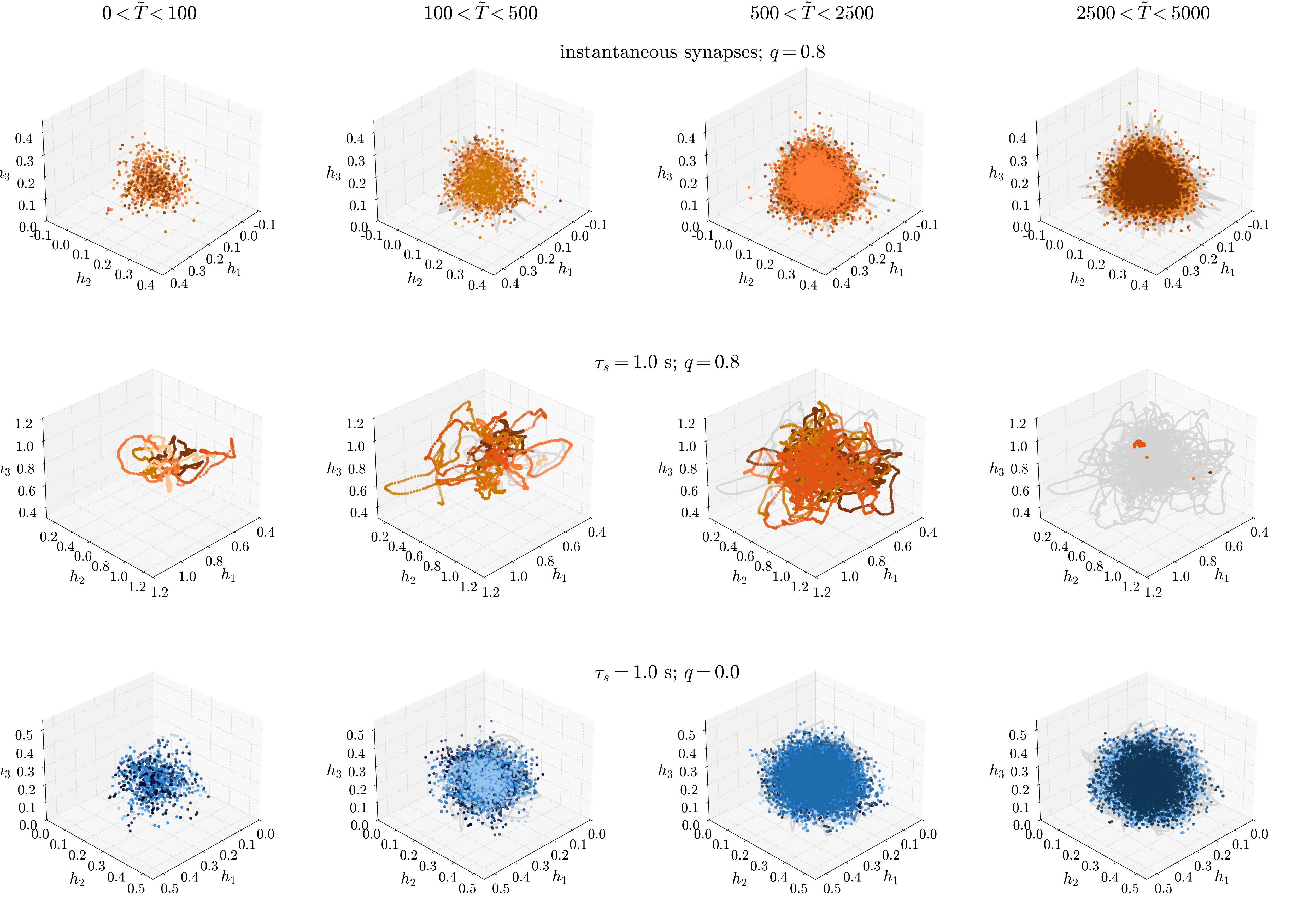}
       \caption{Trajectories in a 3D neural space defined by the total synaptic input of three random neurons; each point corresponds to $(h_1(\tilde{T}), h_2(\tilde{T}), h_3(\tilde{T}))$ at the same time.
       The different colors represent the five different replicas displayed. The upper row corresponds to instantaneous synapses with $q=0.8$, while the other two refer to $\tau_s=1$ s and $q=0.8$ (middle row) and $q=0$ (bottom row). The same parameters as in Fig.\ref{fig:Lif_delta_NetworkActivity} were used.}
    \label{fixedpoints}   
\end{figure*}

It is very insightful to visualize ergodic and non-ergodic dynamics in the network. For this purpose, we use the "projection" of the network dynamics into the three-dimensional space $(h_1, h_2, h_3)$, where $h_i$ denotes the total synaptic input to neuron $i$. In Fig.~\ref{fixedpoints}, we show these projections for five different initial conditions in the same network, with different colors denoting different initial conditions. In the top row, we illustrate the case $q=0.8$ and $\tau_s=0$ (in this case, $\tilde{T}=T$). As can be seen, there is no evident separation between the representative points at any time; the orbits for the different initial conditions fill "densely" the {\em same} volume in the phase space. That this is not an artifact of the low-dimensional projection is, indeed, confirmed by the ergodic distance, $D(\tilde{T})$, that goes to $0$ for $\tilde{T}\to\infty$. For comparison, in the bottom row, we illustrate the case $q=0$ and $\tau_s=1$ s. As is apparent, the qualitative behavior is the same as before. In fact, $D(\tilde{T})\to0$ for $\tilde{T}\to0$ also in the case $q=0$ and $\tau_s=1$ s. By contrast, the behavior is qualitatively different in the case $q=0.8$ and $\tau_s=1$ s, which we illustrate in the middle row.  In this case, while the orbits seem, as before, to fill the same volume at intermediate times, they asymptotically reach clearly separated regions of the phase space. Though it should be evident from the figure (see also Fig.~\ref{fig:ErgodicityBreaking}), it is perhaps worth emphasizing the fact that such "asymptotic confinement" of the orbits---that is, reaching equilibrium---takes very long times, i.e., $\sim10^3 \tau_s$. 

\section{Discussion}
\label{sec:discussion}

We have shown that, in a spiking network operating in the dynamically-balanced regime, the over-expression of reciprocal synaptic connections, naturally and robustly leads to long-lived fluctuations in neural activity. In turn, these fluctuations lead to long relaxation times as estimated by the decay of the autocorrelation function, consistent with experimental observations in the active cortex \cite{ogawa2010differential,murray2014hierarchy,nishida2014discharge}. The emergence of this slow dynamics is independent of both the size of the network and of the synaptic time constant, unlike in previous accounts (as discussed in the Introduction). Rather, it depends on the effective excitatory self-coupling generated by the symmetry in the synaptic connectivity, as already pointed out in \cite{rao2019} and further elaborated upon in the present study. Accordingly, we conclude that long relaxation times are a generic feature of the balanced regime, as soon as one includes in the model a more {\em realistic} description of the local synaptic connectivity \cite{song2005,yoshimura2005,ko2011,brunel2016cortical,campagnola2022local}. 

Besides confirming previous results, here we significantly expand upon them. Specifically, we have shown that the network dynamics becomes non-ergodic when the excitatory self-coupling is suitably strong. Notably, the breaking of ergodicity also occurs in presence of significant levels of fast noise (i.e., for instantaneous synaptic currents) and in partially symmetric networks. By taking long synaptic time constants, we have shown that the network dynamics becomes multi-stable when ergodicity is broken. This suggests that the mechanism underlying the ergodicity breaking in our model network could be a rather classical one, that is, the coexistence of multiple {\em stable} fixed points of the (rate) dynamics. 

It is important to stress that, in our model network, the multi-stability does not result from some low-rank structure embedded in the synaptic connectivity, as is the case, for instance, in model networks of associative memory \cite{hopfield1982neural,amit1997model,curti2004mean} or, more specifically, in clustered networks \cite{deco2012neural,litwin2012,mazzucato2015dynamics}. By contrast, the connectivity matrix in our network has no "structure" beyond a pre-assigned level of symmetry, and it is, indeed, full rank. Therefore, to the best of our knowledge, this is the first report of non-ergodic dynamics in an "unstructured" spiking network (see \cite{berlemont2022glassy} for a related report in a network of rate neurons).  

The usual caveats attached to any result based on numerical simulations apply to our results as well. We have probed the robustness of our results against finite-size effects by systematically varying, within "reasonable" bounds, both the number of neurons in the network ($1000\leq N\leq 5000$) and the average number of connections they make ($100\leq K\leq 3000$). In all cases, we observed the same phenomenology as described in Results; in particular, we found breaking of ergodicity for $q<1$ and $\tau_s=0$, for suitably large values of $g$.

We have also verified that the basic phenomenology does not critically depend on the specific spiking-neuron model. Concretely, we have also studied networks of quadratic integrate-and-fire (QIF) neurons. The QIF neuron is, in some respects, more {\em realistic} than the LIF neuron, since it can approximate reasonably well an entire class of neural models---Type I excitability according to Hodgkin classification---in proximity to the tonic firing threshold \cite{gerstner2014,gutkin2022}. We found, essentially the same phenomenology as for the LIF model. The results for a balanced network of QIF neurons will be presented in a forthcoming publication. 

The best way to assess the robustness of our results would be, obviously, to develop an accompanying analytical theory. This is a formidable task for networks of LIF neurons \cite{lerchner2006response}. On the other hand, using dynamical mean-field theory to achieve an analytical description of the steady states of networks, either of rate elements in presence of white noise or, even better, of spike-response models as in \cite{kadmon2015}, appears feasible although very far from trivial.

Taken together, our results reveal the existence of a novel phase of the collective activity in dynamically-balanced spiking networks. In this phase, the dynamics reaches, asymptotically, different equilibrium states when starting from random initial conditions. In each of these states, the network operates in the balanced regime and, accordingly, the spiking activity of the neurons is unsaturated, temporally-irregular and asynchronous. In this respect, then, it resembles the pattern of activity that one observes in random, ergodic networks. However, unlike in random networks, the ergodicity is broken in this new phase; the memory of the initial conditions never fades away and reaching equilibrium is a very slow process. 

An obvious question, then, is whether there is any evidence that the cortex operates in a non-ergodic phase. We are not aware of any experimental report definitively ruling out this possibility. As just discussed, many features of neural activity will be qualitatively the same in the ergodic and non-ergodic phases. Our work suggests a simple observable, easy to measure in an experiment (at least in principle), that could be help address this question. That is the ergodic distance, $D\left(T\right)$, originally introduced in \cite{berlemont2022glassy}, that we have also used in our spiking network. Note that $D\left(T\right)$ can be computed using only single-cell recordings. Deviations from $1/T$ in its decay at long times (e.g., $\sim 10$s) would be supporting evidence of ergodicity breaking, at least in a weak sense (i.e., very long transients but only one equilibrium state). At present, we are actively pursuing this line of research.  

\section{\label{sec:methods}Methods}

\subsection{\label{sec:model} The model network}

We consider a network of $N$ inhibitory LIF neurons sparsely connected so that each neuron makes and receives, $K$ synaptic contacts on average \cite{monteforte2010,kadmon2015,harish2015asynchronous,di2018}. The neurons also receive a constant excitatory drive from outside the network. The membrane potential of neuron $i$, $v_i$, measured relative to the rest potential and in units of the threshold, evolves in time according to

\begin{equation}\label{eq:LIF_model}
    \tau_m \dot{v}_i = -v_i + h_i(t)
\end{equation}

\noindent where $\tau_m=10$ ms is the membrane time constant and $h_i(t)$ is the total synaptic input (external and recurrent) to neuron $i$. Note that the synaptic input is measured in the same units as the voltage and is thus dimensionless. Whenever $v_i(t) \geq 1$, the voltage is reset to the rest potential $v_i=0$, a spike is generated, and, as a result, postsynaptic currents (PSCs) are initiated at all the postsynaptic targets of neuron $i$. Then, the dynamics Eq.~\eqref{eq:LIF_model} resumes immediately after the spike, starting from $v_i=0$.

The total synaptic input, $h_i(t)$, is given by
\begin{equation}\label{eq:def_h}
    h_i(t)=h_\mathrm{ext}\sqrt{K}-\tau_mE_i(t)
\end{equation}

\noindent where $h_\mathrm{ext}\sqrt{K}>0$ is the external excitatory input and $E_i(t)$ is given by

\begin{equation}\label{eq:LIF_alfa}
    \tau_s \dot{E_i}=-E_i+\frac{g}{\sqrt{K}}\sum_{j=1}^NA_{ij}\sum_{n}\delta(t-t_n^{(j)})
\end{equation}

\noindent where $\tau_s$ is the synaptic time constant, $g>0$ is a measure of the synaptic inhibitory strength, $A_{ij}=1$ if there exists a synaptic connection from neuron $j$ to neuron $i$ ($A_{ij}=0$ otherwise), and the sum over $n$ is over all the spike times, $t_n^{(j)}$, of neuron $j$. In words, $E_i(t)$ is the linear superposition of all the PSCs generated by the spiking activity of the (presynaptic) neurons connected to neuron $i$. The unitary PSC, generated by a single spike (at time $t=0$), is 

\begin{equation}\label{eq:LIF_PSC}
\mathrm{PSC}(t) = \frac{g}{\tau_s\sqrt{K}} {\rm e}^{- t/\tau_s}\Theta(t) 
\end{equation}

\noindent where $\Theta\left(t\right)$ is the Heaviside function. Thus, the synaptic time constant, $\tau_s$, controls the duration of the PSCs. In the limit $\tau_s\to0$, instantaneous synaptic currents, $E_i\left(t\right)$ becomes (see Eq.~\eqref{eq:LIF_alfa})

\begin{equation}\label{eq:LIF_delta}
    E_i=\frac{g}{\sqrt{K}}\sum_{j=1}^NA_{ij}\sum_{n}\delta(t-t_n^{(j)})
\end{equation}

\noindent Note that synaptic currents are measured in units of 1/time (e.g., in Hz).

The unitary PSC, Eq.~\eqref{eq:LIF_PSC}, is proportional to $1/\sqrt{K}$, so that $E_i\left(t\right)\sim\sqrt{K}$ scales with $K$ as the external drive, see Eq.~\eqref{eq:def_h}. To have $h_i\left(t\right)\sim1$ when $K$ is large ($1\ll K\ll N$), the recurrent synaptic input, $\tau_mE_i\left(t\right)$, should cancel (i.e., to leading order) on average the external drive, $h_\mathrm{ext}\sqrt{K}$. Thus, the network operates in the dynamically-balanced regime when $K$ is large \cite{van1996,renart2010asynchronous}. All the simulations described in Results are with $N=4000$, $K=1200$ and $h_\mathrm{ext}=0.1$. 

\subsubsection{\label{sec:network} Synaptic connectivity}

The synaptic connectivity matrix, $\mathbf{A}\equiv\left(A_{ij}\right)$, is generated in the following way \cite{rao2019}. For each {\em ordered} pair of neurons $\left(i,j\right)$, the probability, $p_\mathrm{B}$, that they are reciprocally connected, i.e., $\left(A_{ij}=1,A_{ji}=1\right)$, is given by

\begin{equation}
p_\mathrm{B}=q \cdot \frac{K}{N} + (1 - q)^2 \cdot \frac{K^2}{N^2}    
\end{equation}

\noindent while the probability, $p_\mathrm{U}$, that they are unidirectionally connected, i.e., $\left(A_{ij}=1,A_{ji}=0\right)$, is given by

\begin{equation}
p_\mathrm{U}=(1-q)\cdot \frac{K}{N} - (1-q)^2\cdot \frac{K^2}{N^2}    
\end{equation}

\noindent where the parameter $q$, $0\leq q\leq 1$, controls the over-representation of reciprocal connections, i.e., the level of symmetry. In experiments, this over-representation is often quantified in by the ratio, $\eta$, between the number of {\em observed} reciprocally-connected pairs of neurons divided by the {\em expected} number of reciprocally-connected pairs in a directed Erd\H{o}s-R\'enyi network with the same probability of connection \cite{song2005,campagnola2022local}. The probability of connection is simply estimated by dividing the number of connected pairs by the number of pairs sampled in the experiment. In our case, 

\begin{equation}\label{eq:def_eta}
\eta=\frac{p_\mathrm{B}}{\left(p_\mathrm{B}+p_\mathrm{U}\right)^2}=\frac{q}{K/N}+(1-q)^2    
\end{equation}

\noindent Hence, $\eta=1$ for $q=0$ and the network is simply a directed Erd\H{o}s-R\'enyi network, while $\eta=N/K$ for $q=1$ and all connections are reciprocal, i.e., the network is fully symmetric. 

We note that, while $q$ is independent of the probability of connection (i.e, $K/N$), $\eta$ is not. In particular, the largest possible value of $\eta$ is the reciprocal of the probability of connection. Hence, larger probabilities of connection necessarily result in lower values of $\eta$, regardless of the actual level of symmetry, measured more properly by $q$. For instance, in the cortex local excitatory-to-excitatory connections occur with probability $~\sim 0.1$ and $\eta\simeq 5$ \cite{campagnola2022local}; this leads to $q\simeq0.5$ using Eq.~\eqref{eq:def_eta} (taking, obviously, the solution with $0\leq q\leq 1$). Inhibitory-to-inhibitory connections, instead, occur with probability $~\sim 0.3$ and $\eta\simeq 3$ \cite{campagnola2022local}; this leads to $q\simeq0.9$. Hence, though $\eta$ is larger for the excitatory-to-excitatory network, the inhibitory-to-inhibitory network is more "symmetric". Indeed, about $90\%$ of its connections are bi-directional against only about $50\%$ in the excitatory-to-excitatory network.

\subsubsection{Network simulations}
 
We have employed event-driven techniques to simulate the network dynamics. This allowed us to perform fast and accurate simulations of large networks. In fact, the differential equations, Eqs.\eqref{eq:LIF_model}-\eqref{eq:LIF_alfa}, describing the network dynamics are exactly integrated between consecutive spikes. This can be done by rewriting the evolution of the $v_i$'s and of the $E_i$'s as a mapping between their value at a spiking event occurring at time $t_n$ and their value at the next spiking event at $t_{n+1}$ \cite{zillmer2007}. 

The identity of the neuron firing at the next spiking event and the time, $t_{n+1}$, at which the event occurs can be determined in the following way. Suppose that neuron $k$ has fired at the current spiking event, at time $t_n$. Then, $v_k\left(t_n^{+}\right)=0$ (as a result of the resetting) and

\begin{equation}
    E_i\left(t_n^{+}\right)=E_i\left(t_n^{-}\right)+\frac{g}{\sqrt{K}}\frac{A_{ik}}{\tau_s}
\end{equation}

\noindent where $t_n^{\pm}\equiv t_n\pm\epsilon$ for $\epsilon\to0^{+}$ is the time immediately after/before the spike. After a time interval $\Delta$, if no other spike occurs, we have

\begin{equation}
E_i\left(t_n^{+}+\Delta\right)=E_i\left(t_n^{+}\right)e^{-\Delta/\tau_s}    
\end{equation}

\noindent and

\begin{align}\label{eq:v_recurrent}
v_i\left(t_n^{+}+\Delta\right) & =h_\mathrm{ext}\sqrt{K}+\left(v_i\left(t_n^{+}\right)-h_\mathrm{ext}\sqrt{K}\right)e^{-\Delta/\tau_m}\nonumber \\ & -e^{-\left(t_n^{+}+\Delta\right)/\tau_m}\intop_{t_n^{+}}^{t_n^{+}+\Delta}\mathrm{d}t\;e^{t/\tau_m}E_i\left(t\right)
\end{align}

\noindent The last term on the r.h.s. is 

\begin{equation}\label{eq:integral}
    -E_i\left(t_n^{+}\right)\frac{\tau_m\tau_s}{\tau_m-\tau_s}\left(e^{-\Delta/\tau_m}-e^{-\Delta/\tau_s}\right)
\end{equation}

\noindent Using Eqs.~\eqref{eq:v_recurrent} and \eqref{eq:integral}, we can determine 
the next spiking time, relative to $t_n$, for each neuron $i$ if no other spike occurs before, i.e., $\Delta_i$ such that $v_i\left(t_n^{+}+\Delta_i\right)=1$ for each $i$. Then, the next {\em actual} spiking event occurs at time

\begin{equation}
    t_{n+1}=t_n+\underset{i}{\mathrm{min}}\;\Delta_i
\end{equation}

\noindent and the identity of the neuron spiking at $t_{n+1}$, $m$, is simply 

\begin{equation}
    m =\underset{i}{\mathrm{arg\,min}}\;\Delta_i
\end{equation}

\noindent that is, the neuron $i$ for which $\Delta_i$ achieves its minimum value. 

In the case of instantaneous synapses, we use Eq.~\eqref{eq:LIF_delta} into Eq.~\eqref{eq:v_recurrent}, and the last term on the r.h.s. of Eq.~\eqref{eq:v_recurrent} is now

\begin{equation}
    -\frac{g}{\sqrt{K}}A_{ik}e^{-\Delta/\tau_m}
\end{equation}

\subsection{\label{sec:indicators}Observables}
\subsubsection{Single-neuron firing activity}

The single-neuron activity has been characterized in terms of the following observables: \\

\noindent{\em The firing rate}. For each neuron $i$, we obtained its {\em average} firing rate, $\nu_i$, by counting all the spikes that the neuron fired during the simulation and then dividing the spike count by the total simulation time, $T$. The population-averaged firing rate, $\langle\nu\rangle$, is given by

\begin{equation}
\langle \nu \rangle = \frac{1}{N} \sum_{i=1}^N \nu_i
\end{equation}
    
\noindent Estimates of the $\nu_i$'s were obtained by using simulations with $T=1000$s or longer. In some cases (see, e.g., Fig.~\ref{fig:RatesInReplicas}), we also estimated the {\em instantaneous} firing rate of neuron $i$, $r_i$, that is, the firing rate in a given time window of duration $\Delta<T$. This was obtained by counting only the spikes that the neuron fired {\em within} the time window and then dividing this spike count by $\Delta$. \\

\noindent {\em The coefficient of variation}. The coefficient of variation of neuron $i$, $CV_i$, is the ratio between the standard deviation of its inter-spike intervals and the average of its inter-spike intervals \cite{dayan2001theoretical}. $CV_i$ provides a simple quantification of the degree of temporal irregularity of the spiking process. If neuron $i$ spikes periodically, then $CV_i=0$; if neuron $i$ spikes according to a Poisson process, then $CV_i=1$. The average and standard deviation of the inter-spike intervals have been estimated with simulations of $T=1000$s. 

\subsubsection{Pairwise spike-count correlations}

We choose the bin size $\Delta=100$ ms and divide the interval $\left[0,T\right]$, where $T$ is the total duration of the simulation, into $T/\Delta$ non-overlapping bins. Then, for each neuron $i$ we compute

\begin{equation}
    \delta n_i\left(k\right)=-\nu_i\Delta+\intop_{\left(k-1\right)\Delta}^{k\Delta}\mathrm{d}t\;\sum_{a}\delta(t-t_a^{(i)})
\end{equation}

\noindent where $k=1,\ldots,T/\Delta$ is the bin number, $\nu_i$ is the average firing rate of neuron $i$ estimated in $\left[0,T\right]$, and the sum over $a$ is over all the spike times, $t_a^{(i)}$, of neuron $i$; $\nu_i\Delta$ is the expected spike count in a time bin $\Delta$. Hence, 

\begin{equation}
    \langle\delta n_i\rangle\equiv\frac{\Delta}{T}\sum_{k=1}^{T/\Delta}\delta n_i\left(k\right)=0
\end{equation}

\noindent for each $i=1,\ldots,N$. Then, for each pair of neurons $\left(i,j\right)$, $\rho_{ij}$ is computed as

\begin{equation}
    \rho_{ij}=\rho_{ji}=\frac{\langle\delta n_i\delta n_j\rangle}{\sqrt{\langle\delta n^2_i\rangle\langle\delta n^2_j\rangle}}
\end{equation}

\noindent and it lies between $-1$ and $1$.

\subsubsection{Autocorrelation of the synaptic input}

For each neuron $i$, we compute

\begin{equation}\label{eq:def_delta_h}
\delta h_i\left(t\right)=h_i(t)-\frac{1}{T}\intop_{0}^{T}\mathrm{d}t\;h_i\left(t\right)    
\end{equation}

\noindent where $T$ is the total duration of the simulation. In the case of instantaneous synapses, we low-pass filter $h_i\left(t\right)$ with a cutoff frequency $f_c = 50$Hz before using Eq.~\eqref{eq:def_delta_h}. Hence,

\begin{equation}
    \overline{\delta h_i}\equiv\frac{1}{T}\intop_{0}^{T}\mathrm{d}t\;\delta h_i\left(t\right)=0
\end{equation}

\noindent for each $i=1,\ldots,N$. We define the {\em normalized} autocorrelation, $C_i\left(\tau\right)$, of the input to neuron $i$ at a time lag $\tau$ as

\begin{equation}
    C_i\left(\tau\right)\equiv\frac{1}{\overline{\delta h_i^2}\left(T-\tau\right)}\intop_{0}^{T-\tau}\mathrm{d}t\;\delta h_i\left(t\right)\delta h_i\left(t+\tau\right)
\end{equation}

\noindent so that $C_i\left(0\right)=1$ for each $i=1,\ldots,N$. Finally,

\begin{equation}
    C\left(\tau\right)\equiv\frac{1}{N}\sum_{i=1}^NC_i\left(\tau\right)
\end{equation}

\subsubsection{Spike-triggered input}

For each neuron $i$ and lag $\tau$, we define $Z_i\left(\tau\right)$ as the set of spike numbers, $n$, for which $t_{n}^{(i)}+\tau<t_{n+1}^{(i)}$ if $\tau>0$, or the $n$'s for which $t_{n}^{(i)}+\tau>t_{n-1}^{(i)}$ if $\tau<0$. Thus, if $n\in Z_i\left(\tau\right)$, there are no spikes between $t_{n}^{(i)}$ and $t_{n}^{(i)}+\tau$.

The spike-triggered average input to neuron $i$ at lag $\tau$, $\mathrm{STA}_i\left(\tau\right)$, is given by

\begin{equation}\label{eq:STA_i}
    \mathrm{STA}_i\left(\tau\right)=\frac{1}{\lvert Z_i\left(\tau\right)\rvert}\sum_{n\in Z_i\left(\tau\right)}\left(h_i\left(t_{n}^{(i)}+\tau\right)-\overline{h_i}\right)
\end{equation}

\noindent where $\lvert Z_i\left(\tau\right)\rvert$ is the number of elements in $Z_i\left(\tau\right)$, the sum is restricted to elements in $Z_i\left(\tau\right)$ and $\overline{h_i}$ denotes the average of $h_i\left(t\right)$ over time, i.e.,

\begin{equation}\label{eq:avg_h_over_t}
    \overline{h_i}\equiv\frac{1}{T}\intop_{0}^{T}\mathrm{d}t\;h_i\left(t\right)
\end{equation}

\noindent We low-pass filtered $h_i(t)$ with a cutoff frequency $f_c = 100$ Hz before computing Eqs.~\eqref{eq:STA_i} and \eqref{eq:avg_h_over_t}. By averaging over neurons, we finally obtain the spike-triggered average input, $\mathrm{STA}\left(\tau\right)$, shown in Fig.~\ref{fig:ResponseToInput}, i.e.,

\begin{equation}
    \mathrm{STA}\left(\tau\right)=\frac{1}{N}\sum_{i=1}^N\mathrm{STA}_i\left(\tau\right)
\end{equation}

\subsubsection{Ergodic "distance"}
    
To investigate the ergodicity of the network dynamics, we consider $M$ different {\em physical} replicas of the same network. Each replica is the time evolution of the network starting from different initial conditions. Then, for two different replicas, $\alpha$ and $\beta$, we compute
\begin{eqnarray}\label{eq:Dab}
D^{\alpha\beta}(T) &=& \frac{1}{N}\sum_{i=1}^{N}\left[\frac{n^{\alpha}_i(T)}{T}-\frac{n^{\beta}_i(T)}{T}\right]^2 \\
& = &\frac{1}{N}\sum_{i=1}^{N}\left[\nu^{\alpha}_i(T)-\nu^{\beta}_i(T)\right]^2 \nonumber
\end{eqnarray}
where $n_i^{\alpha}(T)$ is the spike count of the neuron $i$ up to time $T$ in replica $\alpha$ and $\nu_i^\alpha(T)$ is the corresponding firing rate. By averaging over all the distinct pairs of replicas, we finally obtain $D\left(T\right)$ \cite{berlemont2022glassy}, that is,  

\begin{equation}\label{eq:D}
D(T) = \frac{2}{M(M-1)}\sum_{\alpha \neq \beta} D_{\alpha \beta}(T)
\end{equation}
//

\section*{Acknowledgement}
We acknowledge stimulating interactions with Alberto Bacci, Alejandro Carballosa, and Nina La Miciotta. G.M. would like to thank Stefano Fusi for a useful discussion about slow changes in excitability at the single-neuron level. A. T. received financial support by the Labex MME-DII (Grant No. ANR-11-LBX-0023-01) (together with F.T.) and by CY Generations (Grant No ANR-21-EXES-0008) (together with G.M.), all part of the French program Investissements d'Avenir.

\bibliography{references}

\begin{thebibliography}{49}%
\makeatletter
\providecommand \@ifxundefined [1]{%
 \@ifx{#1\undefined}
}%
\providecommand \@ifnum [1]{%
 \ifnum #1\expandafter \@firstoftwo
 \else \expandafter \@secondoftwo
 \fi
}%
\providecommand \@ifx [1]{%
 \ifx #1\expandafter \@firstoftwo
 \else \expandafter \@secondoftwo
 \fi
}%
\providecommand \natexlab [1]{#1}%
\providecommand \enquote  [1]{``#1''}%
\providecommand \bibnamefont  [1]{#1}%
\providecommand \bibfnamefont [1]{#1}%
\providecommand \citenamefont [1]{#1}%
\providecommand \href@noop [0]{\@secondoftwo}%
\providecommand \href [0]{\begingroup \@sanitize@url \@href}%
\providecommand \@href[1]{\@@startlink{#1}\@@href}%
\providecommand \@@href[1]{\endgroup#1\@@endlink}%
\providecommand \@sanitize@url [0]{\catcode `\\12\catcode `\$12\catcode `\&12\catcode `\#12\catcode `\^12\catcode `\_12\catcode `\%12\relax}%
\providecommand \@@startlink[1]{}%
\providecommand \@@endlink[0]{}%
\providecommand \url  [0]{\begingroup\@sanitize@url \@url }%
\providecommand \@url [1]{\endgroup\@href {#1}{\urlprefix }}%
\providecommand \urlprefix  [0]{URL }%
\providecommand \Eprint [0]{\href }%
\providecommand \doibase [0]{http://dx.doi.org/}%
\providecommand \selectlanguage [0]{\@gobble}%
\providecommand \bibinfo  [0]{\@secondoftwo}%
\providecommand \bibfield  [0]{\@secondoftwo}%
\providecommand \translation [1]{[#1]}%
\providecommand \BibitemOpen [0]{}%
\providecommand \bibitemStop [0]{}%
\providecommand \bibitemNoStop [0]{.\EOS\space}%
\providecommand \EOS [0]{\spacefactor3000\relax}%
\providecommand \BibitemShut  [1]{\csname bibitem#1\endcsname}%
\let\auto@bib@innerbib\@empty
\bibitem [{\citenamefont {Ogawa}\ and\ \citenamefont {Komatsu}(2010)}]{ogawa2010differential}%
  \BibitemOpen
  \bibfield  {author} {\bibinfo {author} {\bibfnamefont {T.}~\bibnamefont {Ogawa}}\ and\ \bibinfo {author} {\bibfnamefont {H.}~\bibnamefont {Komatsu}},\ }\href@noop {} {\bibfield  {journal} {\bibinfo  {journal} {Journal of neurophysiology}\ }\textbf {\bibinfo {volume} {103}},\ \bibinfo {pages} {2433} (\bibinfo {year} {2010})}\BibitemShut {NoStop}%
\bibitem [{\citenamefont {Murray}\ \emph {et~al.}(2014)\citenamefont {Murray}, \citenamefont {Bernacchia}, \citenamefont {Freedman}, \citenamefont {Romo}, \citenamefont {Wallis}, \citenamefont {Cai}, \citenamefont {Padoa-Schioppa}, \citenamefont {Pasternak}, \citenamefont {Seo}, \citenamefont {Lee} \emph {et~al.}}]{murray2014hierarchy}%
  \BibitemOpen
  \bibfield  {author} {\bibinfo {author} {\bibfnamefont {J.~D.}\ \bibnamefont {Murray}}, \bibinfo {author} {\bibfnamefont {A.}~\bibnamefont {Bernacchia}}, \bibinfo {author} {\bibfnamefont {D.~J.}\ \bibnamefont {Freedman}}, \bibinfo {author} {\bibfnamefont {R.}~\bibnamefont {Romo}}, \bibinfo {author} {\bibfnamefont {J.~D.}\ \bibnamefont {Wallis}}, \bibinfo {author} {\bibfnamefont {X.}~\bibnamefont {Cai}}, \bibinfo {author} {\bibfnamefont {C.}~\bibnamefont {Padoa-Schioppa}}, \bibinfo {author} {\bibfnamefont {T.}~\bibnamefont {Pasternak}}, \bibinfo {author} {\bibfnamefont {H.}~\bibnamefont {Seo}}, \bibinfo {author} {\bibfnamefont {D.}~\bibnamefont {Lee}},  \emph {et~al.},\ }\href@noop {} {\bibfield  {journal} {\bibinfo  {journal} {Nature neuroscience}\ }\textbf {\bibinfo {volume} {17}},\ \bibinfo {pages} {1661} (\bibinfo {year} {2014})}\BibitemShut {NoStop}%
\bibitem [{\citenamefont {Nishida}\ \emph {et~al.}(2014)\citenamefont {Nishida}, \citenamefont {Tanaka}, \citenamefont {Shibata}, \citenamefont {Ikeda}, \citenamefont {Aso},\ and\ \citenamefont {Ogawa}}]{nishida2014discharge}%
  \BibitemOpen
  \bibfield  {author} {\bibinfo {author} {\bibfnamefont {S.}~\bibnamefont {Nishida}}, \bibinfo {author} {\bibfnamefont {T.}~\bibnamefont {Tanaka}}, \bibinfo {author} {\bibfnamefont {T.}~\bibnamefont {Shibata}}, \bibinfo {author} {\bibfnamefont {K.}~\bibnamefont {Ikeda}}, \bibinfo {author} {\bibfnamefont {T.}~\bibnamefont {Aso}}, \ and\ \bibinfo {author} {\bibfnamefont {T.}~\bibnamefont {Ogawa}},\ }\href@noop {} {\bibfield  {journal} {\bibinfo  {journal} {Cerebral Cortex}\ }\textbf {\bibinfo {volume} {24}},\ \bibinfo {pages} {1671} (\bibinfo {year} {2014})}\BibitemShut {NoStop}%
\bibitem [{\citenamefont {Dayan}\ and\ \citenamefont {Abott}(2001)}]{dayan2001theoretical}%
  \BibitemOpen
  \bibfield  {author} {\bibinfo {author} {\bibfnamefont {P.}~\bibnamefont {Dayan}}\ and\ \bibinfo {author} {\bibfnamefont {L.}~\bibnamefont {Abott}},\ }\href@noop {} {\enquote {\bibinfo {title} {Theoretical neuroscience},}\ } (\bibinfo {year} {2001})\BibitemShut {NoStop}%
\bibitem [{\citenamefont {La~Camera}\ \emph {et~al.}(2006)\citenamefont {La~Camera}, \citenamefont {Rauch}, \citenamefont {Thurbon}, \citenamefont {Luscher}, \citenamefont {Senn},\ and\ \citenamefont {Fusi}}]{la2006multiple}%
  \BibitemOpen
  \bibfield  {author} {\bibinfo {author} {\bibfnamefont {G.}~\bibnamefont {La~Camera}}, \bibinfo {author} {\bibfnamefont {A.}~\bibnamefont {Rauch}}, \bibinfo {author} {\bibfnamefont {D.}~\bibnamefont {Thurbon}}, \bibinfo {author} {\bibfnamefont {H.-R.}\ \bibnamefont {Luscher}}, \bibinfo {author} {\bibfnamefont {W.}~\bibnamefont {Senn}}, \ and\ \bibinfo {author} {\bibfnamefont {S.}~\bibnamefont {Fusi}},\ }\href@noop {} {\bibfield  {journal} {\bibinfo  {journal} {Journal of neurophysiology}\ }\textbf {\bibinfo {volume} {96}},\ \bibinfo {pages} {3448} (\bibinfo {year} {2006})}\BibitemShut {NoStop}%
\bibitem [{\citenamefont {Pozzorini}\ \emph {et~al.}(2013)\citenamefont {Pozzorini}, \citenamefont {Naud}, \citenamefont {Mensi},\ and\ \citenamefont {Gerstner}}]{pozzorini2013temporal}%
  \BibitemOpen
  \bibfield  {author} {\bibinfo {author} {\bibfnamefont {C.}~\bibnamefont {Pozzorini}}, \bibinfo {author} {\bibfnamefont {R.}~\bibnamefont {Naud}}, \bibinfo {author} {\bibfnamefont {S.}~\bibnamefont {Mensi}}, \ and\ \bibinfo {author} {\bibfnamefont {W.}~\bibnamefont {Gerstner}},\ }\href@noop {} {\bibfield  {journal} {\bibinfo  {journal} {Nature neuroscience}\ }\textbf {\bibinfo {volume} {16}},\ \bibinfo {pages} {942} (\bibinfo {year} {2013})}\BibitemShut {NoStop}%
\bibitem [{\citenamefont {Fisher}\ \emph {et~al.}(1997)\citenamefont {Fisher}, \citenamefont {Fischer},\ and\ \citenamefont {Carew}}]{fisher1997multiple}%
  \BibitemOpen
  \bibfield  {author} {\bibinfo {author} {\bibfnamefont {S.~A.}\ \bibnamefont {Fisher}}, \bibinfo {author} {\bibfnamefont {T.~M.}\ \bibnamefont {Fischer}}, \ and\ \bibinfo {author} {\bibfnamefont {T.~J.}\ \bibnamefont {Carew}},\ }\href@noop {} {\bibfield  {journal} {\bibinfo  {journal} {Trends in neurosciences}\ }\textbf {\bibinfo {volume} {20}},\ \bibinfo {pages} {170} (\bibinfo {year} {1997})}\BibitemShut {NoStop}%
\bibitem [{\citenamefont {Beiran}\ and\ \citenamefont {Ostojic}(2019)}]{beiran2019}%
  \BibitemOpen
  \bibfield  {author} {\bibinfo {author} {\bibfnamefont {M.}~\bibnamefont {Beiran}}\ and\ \bibinfo {author} {\bibfnamefont {S.}~\bibnamefont {Ostojic}},\ }\href@noop {} {\bibfield  {journal} {\bibinfo  {journal} {PLoS computational biology}\ }\textbf {\bibinfo {volume} {15}},\ \bibinfo {pages} {e1006893} (\bibinfo {year} {2019})}\BibitemShut {NoStop}%
\bibitem [{\citenamefont {Tsodyks}\ and\ \citenamefont {Sejnowski}(1995)}]{tsodyks1995rapid}%
  \BibitemOpen
  \bibfield  {author} {\bibinfo {author} {\bibfnamefont {M.~V.}\ \bibnamefont {Tsodyks}}\ and\ \bibinfo {author} {\bibfnamefont {T.}~\bibnamefont {Sejnowski}},\ }\href@noop {} {\bibfield  {journal} {\bibinfo  {journal} {Network: Computation in Neural Systems}\ }\textbf {\bibinfo {volume} {6}},\ \bibinfo {pages} {111} (\bibinfo {year} {1995})}\BibitemShut {NoStop}%
\bibitem [{\citenamefont {Van~Vreeswijk}\ and\ \citenamefont {Sompolinsky}(1996)}]{van1996}%
  \BibitemOpen
  \bibfield  {author} {\bibinfo {author} {\bibfnamefont {C.}~\bibnamefont {Van~Vreeswijk}}\ and\ \bibinfo {author} {\bibfnamefont {H.}~\bibnamefont {Sompolinsky}},\ }\href@noop {} {\bibfield  {journal} {\bibinfo  {journal} {Science}\ }\textbf {\bibinfo {volume} {274}},\ \bibinfo {pages} {1724} (\bibinfo {year} {1996})}\BibitemShut {NoStop}%
\bibitem [{\citenamefont {Amit}\ and\ \citenamefont {Brunel}(1997)}]{amit1997model}%
  \BibitemOpen
  \bibfield  {author} {\bibinfo {author} {\bibfnamefont {D.~J.}\ \bibnamefont {Amit}}\ and\ \bibinfo {author} {\bibfnamefont {N.}~\bibnamefont {Brunel}},\ }\href@noop {} {\bibfield  {journal} {\bibinfo  {journal} {Cerebral cortex (New York, NY: 1991)}\ }\textbf {\bibinfo {volume} {7}},\ \bibinfo {pages} {237} (\bibinfo {year} {1997})}\BibitemShut {NoStop}%
\bibitem [{\citenamefont {Renart}\ \emph {et~al.}(2010)\citenamefont {Renart}, \citenamefont {De~La~Rocha}, \citenamefont {Bartho}, \citenamefont {Hollender}, \citenamefont {Parga}, \citenamefont {Reyes},\ and\ \citenamefont {Harris}}]{renart2010asynchronous}%
  \BibitemOpen
  \bibfield  {author} {\bibinfo {author} {\bibfnamefont {A.}~\bibnamefont {Renart}}, \bibinfo {author} {\bibfnamefont {J.}~\bibnamefont {De~La~Rocha}}, \bibinfo {author} {\bibfnamefont {P.}~\bibnamefont {Bartho}}, \bibinfo {author} {\bibfnamefont {L.}~\bibnamefont {Hollender}}, \bibinfo {author} {\bibfnamefont {N.}~\bibnamefont {Parga}}, \bibinfo {author} {\bibfnamefont {A.}~\bibnamefont {Reyes}}, \ and\ \bibinfo {author} {\bibfnamefont {K.~D.}\ \bibnamefont {Harris}},\ }\href@noop {} {\bibfield  {journal} {\bibinfo  {journal} {science}\ }\textbf {\bibinfo {volume} {327}},\ \bibinfo {pages} {587} (\bibinfo {year} {2010})}\BibitemShut {NoStop}%
\bibitem [{\citenamefont {Lerchner}\ \emph {et~al.}(2006)\citenamefont {Lerchner}, \citenamefont {Ursta}, \citenamefont {Hertz}, \citenamefont {Ahmadi}, \citenamefont {Ruffiot},\ and\ \citenamefont {Enemark}}]{lerchner2006response}%
  \BibitemOpen
  \bibfield  {author} {\bibinfo {author} {\bibfnamefont {A.}~\bibnamefont {Lerchner}}, \bibinfo {author} {\bibfnamefont {C.}~\bibnamefont {Ursta}}, \bibinfo {author} {\bibfnamefont {J.}~\bibnamefont {Hertz}}, \bibinfo {author} {\bibfnamefont {M.}~\bibnamefont {Ahmadi}}, \bibinfo {author} {\bibfnamefont {P.}~\bibnamefont {Ruffiot}}, \ and\ \bibinfo {author} {\bibfnamefont {S.}~\bibnamefont {Enemark}},\ }\href@noop {} {\bibfield  {journal} {\bibinfo  {journal} {Neural computation}\ }\textbf {\bibinfo {volume} {18}},\ \bibinfo {pages} {634} (\bibinfo {year} {2006})}\BibitemShut {NoStop}%
\bibitem [{\citenamefont {Roxin}\ \emph {et~al.}(2011)\citenamefont {Roxin}, \citenamefont {Brunel}, \citenamefont {Hansel}, \citenamefont {Mongillo},\ and\ \citenamefont {van Vreeswijk}}]{roxin2011distribution}%
  \BibitemOpen
  \bibfield  {author} {\bibinfo {author} {\bibfnamefont {A.}~\bibnamefont {Roxin}}, \bibinfo {author} {\bibfnamefont {N.}~\bibnamefont {Brunel}}, \bibinfo {author} {\bibfnamefont {D.}~\bibnamefont {Hansel}}, \bibinfo {author} {\bibfnamefont {G.}~\bibnamefont {Mongillo}}, \ and\ \bibinfo {author} {\bibfnamefont {C.}~\bibnamefont {van Vreeswijk}},\ }\href@noop {} {\bibfield  {journal} {\bibinfo  {journal} {Journal of Neuroscience}\ }\textbf {\bibinfo {volume} {31}},\ \bibinfo {pages} {16217} (\bibinfo {year} {2011})}\BibitemShut {NoStop}%
\bibitem [{\citenamefont {Mongillo}\ \emph {et~al.}(2012)\citenamefont {Mongillo}, \citenamefont {Hansel},\ and\ \citenamefont {Van~Vreeswijk}}]{mongillo2012bistability}%
  \BibitemOpen
  \bibfield  {author} {\bibinfo {author} {\bibfnamefont {G.}~\bibnamefont {Mongillo}}, \bibinfo {author} {\bibfnamefont {D.}~\bibnamefont {Hansel}}, \ and\ \bibinfo {author} {\bibfnamefont {C.}~\bibnamefont {Van~Vreeswijk}},\ }\href@noop {} {\bibfield  {journal} {\bibinfo  {journal} {Physical review letters}\ }\textbf {\bibinfo {volume} {108}},\ \bibinfo {pages} {158101} (\bibinfo {year} {2012})}\BibitemShut {NoStop}%
\bibitem [{\citenamefont {Mongillo}\ \emph {et~al.}(2018)\citenamefont {Mongillo}, \citenamefont {Rumpel},\ and\ \citenamefont {Loewenstein}}]{mongillo2018inhibitory}%
  \BibitemOpen
  \bibfield  {author} {\bibinfo {author} {\bibfnamefont {G.}~\bibnamefont {Mongillo}}, \bibinfo {author} {\bibfnamefont {S.}~\bibnamefont {Rumpel}}, \ and\ \bibinfo {author} {\bibfnamefont {Y.}~\bibnamefont {Loewenstein}},\ }\href@noop {} {\bibfield  {journal} {\bibinfo  {journal} {Nature neuroscience}\ }\textbf {\bibinfo {volume} {21}},\ \bibinfo {pages} {1463} (\bibinfo {year} {2018})}\BibitemShut {NoStop}%
\bibitem [{\citenamefont {Helias}\ \emph {et~al.}(2014)\citenamefont {Helias}, \citenamefont {Tetzlaff},\ and\ \citenamefont {Diesmann}}]{helias2014correlation}%
  \BibitemOpen
  \bibfield  {author} {\bibinfo {author} {\bibfnamefont {M.}~\bibnamefont {Helias}}, \bibinfo {author} {\bibfnamefont {T.}~\bibnamefont {Tetzlaff}}, \ and\ \bibinfo {author} {\bibfnamefont {M.}~\bibnamefont {Diesmann}},\ }\href@noop {} {\bibfield  {journal} {\bibinfo  {journal} {PLoS computational biology}\ }\textbf {\bibinfo {volume} {10}},\ \bibinfo {pages} {e1003428} (\bibinfo {year} {2014})}\BibitemShut {NoStop}%
\bibitem [{\citenamefont {Harish}\ and\ \citenamefont {Hansel}(2015)}]{harish2015asynchronous}%
  \BibitemOpen
  \bibfield  {author} {\bibinfo {author} {\bibfnamefont {O.}~\bibnamefont {Harish}}\ and\ \bibinfo {author} {\bibfnamefont {D.}~\bibnamefont {Hansel}},\ }\href@noop {} {\bibfield  {journal} {\bibinfo  {journal} {PLoS computational biology}\ }\textbf {\bibinfo {volume} {11}},\ \bibinfo {pages} {e1004266} (\bibinfo {year} {2015})}\BibitemShut {NoStop}%
\bibitem [{\citenamefont {Kadmon}\ and\ \citenamefont {Sompolinsky}(2015)}]{kadmon2015}%
  \BibitemOpen
  \bibfield  {author} {\bibinfo {author} {\bibfnamefont {J.}~\bibnamefont {Kadmon}}\ and\ \bibinfo {author} {\bibfnamefont {H.}~\bibnamefont {Sompolinsky}},\ }\href@noop {} {\bibfield  {journal} {\bibinfo  {journal} {Physical Review X}\ }\textbf {\bibinfo {volume} {5}},\ \bibinfo {pages} {041030} (\bibinfo {year} {2015})}\BibitemShut {NoStop}%
\bibitem [{\citenamefont {Dahmen}\ \emph {et~al.}(2019)\citenamefont {Dahmen}, \citenamefont {Gr{\"u}n}, \citenamefont {Diesmann},\ and\ \citenamefont {Helias}}]{dahmen2019second}%
  \BibitemOpen
  \bibfield  {author} {\bibinfo {author} {\bibfnamefont {D.}~\bibnamefont {Dahmen}}, \bibinfo {author} {\bibfnamefont {S.}~\bibnamefont {Gr{\"u}n}}, \bibinfo {author} {\bibfnamefont {M.}~\bibnamefont {Diesmann}}, \ and\ \bibinfo {author} {\bibfnamefont {M.}~\bibnamefont {Helias}},\ }\href@noop {} {\bibfield  {journal} {\bibinfo  {journal} {Proceedings of the National Academy of Sciences}\ }\textbf {\bibinfo {volume} {116}},\ \bibinfo {pages} {13051} (\bibinfo {year} {2019})}\BibitemShut {NoStop}%
\bibitem [{\citenamefont {Angulo-Garcia}\ \emph {et~al.}(2017)\citenamefont {Angulo-Garcia}, \citenamefont {Luccioli}, \citenamefont {Olmi},\ and\ \citenamefont {Torcini}}]{angulo2017}%
  \BibitemOpen
  \bibfield  {author} {\bibinfo {author} {\bibfnamefont {D.}~\bibnamefont {Angulo-Garcia}}, \bibinfo {author} {\bibfnamefont {S.}~\bibnamefont {Luccioli}}, \bibinfo {author} {\bibfnamefont {S.}~\bibnamefont {Olmi}}, \ and\ \bibinfo {author} {\bibfnamefont {A.}~\bibnamefont {Torcini}},\ }\href@noop {} {\bibfield  {journal} {\bibinfo  {journal} {New Journal of Physics}\ }\textbf {\bibinfo {volume} {19}},\ \bibinfo {pages} {053011} (\bibinfo {year} {2017})}\BibitemShut {NoStop}%
\bibitem [{\citenamefont {Deco}\ and\ \citenamefont {Hugues}(2012)}]{deco2012neural}%
  \BibitemOpen
  \bibfield  {author} {\bibinfo {author} {\bibfnamefont {G.}~\bibnamefont {Deco}}\ and\ \bibinfo {author} {\bibfnamefont {E.}~\bibnamefont {Hugues}},\ }\href@noop {} {\bibfield  {journal} {\bibinfo  {journal} {PLoS computational biology}\ }\textbf {\bibinfo {volume} {8}},\ \bibinfo {pages} {e1002395} (\bibinfo {year} {2012})}\BibitemShut {NoStop}%
\bibitem [{\citenamefont {Litwin-Kumar}\ and\ \citenamefont {Doiron}(2012)}]{litwin2012}%
  \BibitemOpen
  \bibfield  {author} {\bibinfo {author} {\bibfnamefont {A.}~\bibnamefont {Litwin-Kumar}}\ and\ \bibinfo {author} {\bibfnamefont {B.}~\bibnamefont {Doiron}},\ }\href@noop {} {\bibfield  {journal} {\bibinfo  {journal} {Nature neuroscience}\ }\textbf {\bibinfo {volume} {15}},\ \bibinfo {pages} {1498} (\bibinfo {year} {2012})}\BibitemShut {NoStop}%
\bibitem [{\citenamefont {Mazzucato}\ \emph {et~al.}(2015)\citenamefont {Mazzucato}, \citenamefont {Fontanini},\ and\ \citenamefont {La~Camera}}]{mazzucato2015dynamics}%
  \BibitemOpen
  \bibfield  {author} {\bibinfo {author} {\bibfnamefont {L.}~\bibnamefont {Mazzucato}}, \bibinfo {author} {\bibfnamefont {A.}~\bibnamefont {Fontanini}}, \ and\ \bibinfo {author} {\bibfnamefont {G.}~\bibnamefont {La~Camera}},\ }\href@noop {} {\bibfield  {journal} {\bibinfo  {journal} {Journal of Neuroscience}\ }\textbf {\bibinfo {volume} {35}},\ \bibinfo {pages} {8214} (\bibinfo {year} {2015})}\BibitemShut {NoStop}%
\bibitem [{\citenamefont {Brinkman}\ \emph {et~al.}(2022)\citenamefont {Brinkman}, \citenamefont {Yan}, \citenamefont {Maffei}, \citenamefont {Park}, \citenamefont {Fontanini}, \citenamefont {Wang},\ and\ \citenamefont {La~Camera}}]{brinkman2022metastable}%
  \BibitemOpen
  \bibfield  {author} {\bibinfo {author} {\bibfnamefont {B.~A.}\ \bibnamefont {Brinkman}}, \bibinfo {author} {\bibfnamefont {H.}~\bibnamefont {Yan}}, \bibinfo {author} {\bibfnamefont {A.}~\bibnamefont {Maffei}}, \bibinfo {author} {\bibfnamefont {I.~M.}\ \bibnamefont {Park}}, \bibinfo {author} {\bibfnamefont {A.}~\bibnamefont {Fontanini}}, \bibinfo {author} {\bibfnamefont {J.}~\bibnamefont {Wang}}, \ and\ \bibinfo {author} {\bibfnamefont {G.}~\bibnamefont {La~Camera}},\ }\href@noop {} {\bibfield  {journal} {\bibinfo  {journal} {Applied Physics Reviews}\ }\textbf {\bibinfo {volume} {9}} (\bibinfo {year} {2022})}\BibitemShut {NoStop}%
\bibitem [{\citenamefont {Huang}\ and\ \citenamefont {Doiron}(2017)}]{huang2017once}%
  \BibitemOpen
  \bibfield  {author} {\bibinfo {author} {\bibfnamefont {C.}~\bibnamefont {Huang}}\ and\ \bibinfo {author} {\bibfnamefont {B.}~\bibnamefont {Doiron}},\ }\href@noop {} {\bibfield  {journal} {\bibinfo  {journal} {Current opinion in neurobiology}\ }\textbf {\bibinfo {volume} {46}},\ \bibinfo {pages} {31} (\bibinfo {year} {2017})}\BibitemShut {NoStop}%
\bibitem [{\citenamefont {H{\"a}nggi}\ \emph {et~al.}(1990)\citenamefont {H{\"a}nggi}, \citenamefont {Talkner},\ and\ \citenamefont {Borkovec}}]{hanggi1990}%
  \BibitemOpen
  \bibfield  {author} {\bibinfo {author} {\bibfnamefont {P.}~\bibnamefont {H{\"a}nggi}}, \bibinfo {author} {\bibfnamefont {P.}~\bibnamefont {Talkner}}, \ and\ \bibinfo {author} {\bibfnamefont {M.}~\bibnamefont {Borkovec}},\ }\href@noop {} {\bibfield  {journal} {\bibinfo  {journal} {Reviews of modern physics}\ }\textbf {\bibinfo {volume} {62}},\ \bibinfo {pages} {251} (\bibinfo {year} {1990})}\BibitemShut {NoStop}%
\bibitem [{\citenamefont {Yang}\ and\ \citenamefont {La~Camera}(2024)}]{yang2024co}%
  \BibitemOpen
  \bibfield  {author} {\bibinfo {author} {\bibfnamefont {X.}~\bibnamefont {Yang}}\ and\ \bibinfo {author} {\bibfnamefont {G.}~\bibnamefont {La~Camera}},\ }\href@noop {} {\bibfield  {journal} {\bibinfo  {journal} {PLOS Computational Biology}\ }\textbf {\bibinfo {volume} {20}},\ \bibinfo {pages} {e1012220} (\bibinfo {year} {2024})}\BibitemShut {NoStop}%
\bibitem [{\citenamefont {Mart{\'\i}}\ \emph {et~al.}(2018)\citenamefont {Mart{\'\i}}, \citenamefont {Brunel},\ and\ \citenamefont {Ostojic}}]{marti2018}%
  \BibitemOpen
  \bibfield  {author} {\bibinfo {author} {\bibfnamefont {D.}~\bibnamefont {Mart{\'\i}}}, \bibinfo {author} {\bibfnamefont {N.}~\bibnamefont {Brunel}}, \ and\ \bibinfo {author} {\bibfnamefont {S.}~\bibnamefont {Ostojic}},\ }\href@noop {} {\bibfield  {journal} {\bibinfo  {journal} {Physical Review E}\ }\textbf {\bibinfo {volume} {97}},\ \bibinfo {pages} {062314} (\bibinfo {year} {2018})}\BibitemShut {NoStop}%
\bibitem [{\citenamefont {Rao}\ \emph {et~al.}(2019)\citenamefont {Rao}, \citenamefont {Hansel},\ and\ \citenamefont {van Vreeswijk}}]{rao2019}%
  \BibitemOpen
  \bibfield  {author} {\bibinfo {author} {\bibfnamefont {S.}~\bibnamefont {Rao}}, \bibinfo {author} {\bibfnamefont {D.}~\bibnamefont {Hansel}}, \ and\ \bibinfo {author} {\bibfnamefont {C.}~\bibnamefont {van Vreeswijk}},\ }\href@noop {} {\bibfield  {journal} {\bibinfo  {journal} {Scientific Reports}\ }\textbf {\bibinfo {volume} {9}},\ \bibinfo {pages} {3334} (\bibinfo {year} {2019})}\BibitemShut {NoStop}%
\bibitem [{\citenamefont {Berlemont}\ and\ \citenamefont {Mongillo}(2022)}]{berlemont2022glassy}%
  \BibitemOpen
  \bibfield  {author} {\bibinfo {author} {\bibfnamefont {K.}~\bibnamefont {Berlemont}}\ and\ \bibinfo {author} {\bibfnamefont {G.}~\bibnamefont {Mongillo}},\ }\href@noop {} {\bibfield  {journal} {\bibinfo  {journal} {bioRxiv}\ ,\ \bibinfo {pages} {2022}} (\bibinfo {year} {2022})}\BibitemShut {NoStop}%
\bibitem [{\citenamefont {Song}\ \emph {et~al.}(2005)\citenamefont {Song}, \citenamefont {Sj{\"o}str{\"o}m}, \citenamefont {Reigl}, \citenamefont {Nelson},\ and\ \citenamefont {Chklovskii}}]{song2005}%
  \BibitemOpen
  \bibfield  {author} {\bibinfo {author} {\bibfnamefont {S.}~\bibnamefont {Song}}, \bibinfo {author} {\bibfnamefont {P.~J.}\ \bibnamefont {Sj{\"o}str{\"o}m}}, \bibinfo {author} {\bibfnamefont {M.}~\bibnamefont {Reigl}}, \bibinfo {author} {\bibfnamefont {S.}~\bibnamefont {Nelson}}, \ and\ \bibinfo {author} {\bibfnamefont {D.~B.}\ \bibnamefont {Chklovskii}},\ }\href@noop {} {\bibfield  {journal} {\bibinfo  {journal} {PLoS biology}\ }\textbf {\bibinfo {volume} {3}},\ \bibinfo {pages} {e68} (\bibinfo {year} {2005})}\BibitemShut {NoStop}%
\bibitem [{\citenamefont {Yoshimura}\ \emph {et~al.}(2005)\citenamefont {Yoshimura}, \citenamefont {Dantzker},\ and\ \citenamefont {Callaway}}]{yoshimura2005}%
  \BibitemOpen
  \bibfield  {author} {\bibinfo {author} {\bibfnamefont {Y.}~\bibnamefont {Yoshimura}}, \bibinfo {author} {\bibfnamefont {J.~L.}\ \bibnamefont {Dantzker}}, \ and\ \bibinfo {author} {\bibfnamefont {E.~M.}\ \bibnamefont {Callaway}},\ }\href@noop {} {\bibfield  {journal} {\bibinfo  {journal} {Nature}\ }\textbf {\bibinfo {volume} {433}},\ \bibinfo {pages} {868} (\bibinfo {year} {2005})}\BibitemShut {NoStop}%
\bibitem [{\citenamefont {Ko}\ \emph {et~al.}(2011)\citenamefont {Ko}, \citenamefont {Hofer}, \citenamefont {Pichler}, \citenamefont {Buchanan}, \citenamefont {Sj{\"o}str{\"o}m},\ and\ \citenamefont {Mrsic-Flogel}}]{ko2011}%
  \BibitemOpen
  \bibfield  {author} {\bibinfo {author} {\bibfnamefont {H.}~\bibnamefont {Ko}}, \bibinfo {author} {\bibfnamefont {S.~B.}\ \bibnamefont {Hofer}}, \bibinfo {author} {\bibfnamefont {B.}~\bibnamefont {Pichler}}, \bibinfo {author} {\bibfnamefont {K.~A.}\ \bibnamefont {Buchanan}}, \bibinfo {author} {\bibfnamefont {P.~J.}\ \bibnamefont {Sj{\"o}str{\"o}m}}, \ and\ \bibinfo {author} {\bibfnamefont {T.~D.}\ \bibnamefont {Mrsic-Flogel}},\ }\href@noop {} {\bibfield  {journal} {\bibinfo  {journal} {Nature}\ }\textbf {\bibinfo {volume} {473}},\ \bibinfo {pages} {87} (\bibinfo {year} {2011})}\BibitemShut {NoStop}%
\bibitem [{\citenamefont {Brunel}(2016)}]{brunel2016cortical}%
  \BibitemOpen
  \bibfield  {author} {\bibinfo {author} {\bibfnamefont {N.}~\bibnamefont {Brunel}},\ }\href@noop {} {\bibfield  {journal} {\bibinfo  {journal} {Nature neuroscience}\ }\textbf {\bibinfo {volume} {19}},\ \bibinfo {pages} {749} (\bibinfo {year} {2016})}\BibitemShut {NoStop}%
\bibitem [{\citenamefont {Campagnola}\ \emph {et~al.}(2022)\citenamefont {Campagnola}, \citenamefont {Seeman}, \citenamefont {Chartrand}, \citenamefont {Kim}, \citenamefont {Hoggarth}, \citenamefont {Gamlin}, \citenamefont {Ito}, \citenamefont {Trinh}, \citenamefont {Davoudian}, \citenamefont {Radaelli} \emph {et~al.}}]{campagnola2022local}%
  \BibitemOpen
  \bibfield  {author} {\bibinfo {author} {\bibfnamefont {L.}~\bibnamefont {Campagnola}}, \bibinfo {author} {\bibfnamefont {S.~C.}\ \bibnamefont {Seeman}}, \bibinfo {author} {\bibfnamefont {T.}~\bibnamefont {Chartrand}}, \bibinfo {author} {\bibfnamefont {L.}~\bibnamefont {Kim}}, \bibinfo {author} {\bibfnamefont {A.}~\bibnamefont {Hoggarth}}, \bibinfo {author} {\bibfnamefont {C.}~\bibnamefont {Gamlin}}, \bibinfo {author} {\bibfnamefont {S.}~\bibnamefont {Ito}}, \bibinfo {author} {\bibfnamefont {J.}~\bibnamefont {Trinh}}, \bibinfo {author} {\bibfnamefont {P.}~\bibnamefont {Davoudian}}, \bibinfo {author} {\bibfnamefont {C.}~\bibnamefont {Radaelli}},  \emph {et~al.},\ }\href@noop {} {\bibfield  {journal} {\bibinfo  {journal} {Science}\ }\textbf {\bibinfo {volume} {375}},\ \bibinfo {pages} {eabj5861} (\bibinfo {year} {2022})}\BibitemShut {NoStop}%
\bibitem [{\citenamefont {Crisanti}\ and\ \citenamefont {Sompolinsky}(1988)}]{crisanti1988dynamics}%
  \BibitemOpen
  \bibfield  {author} {\bibinfo {author} {\bibfnamefont {A.}~\bibnamefont {Crisanti}}\ and\ \bibinfo {author} {\bibfnamefont {H.}~\bibnamefont {Sompolinsky}},\ }\href@noop {} {\bibfield  {journal} {\bibinfo  {journal} {Physical Review A}\ }\textbf {\bibinfo {volume} {37}},\ \bibinfo {pages} {4865} (\bibinfo {year} {1988})}\BibitemShut {NoStop}%
\bibitem [{\citenamefont {Monteforte}\ and\ \citenamefont {Wolf}(2010)}]{monteforte2010}%
  \BibitemOpen
  \bibfield  {author} {\bibinfo {author} {\bibfnamefont {M.}~\bibnamefont {Monteforte}}\ and\ \bibinfo {author} {\bibfnamefont {F.}~\bibnamefont {Wolf}},\ }\href@noop {} {\bibfield  {journal} {\bibinfo  {journal} {Physical review letters}\ }\textbf {\bibinfo {volume} {105}},\ \bibinfo {pages} {268104} (\bibinfo {year} {2010})}\BibitemShut {NoStop}%
\bibitem [{\citenamefont {Di~Volo}\ and\ \citenamefont {Torcini}(2018)}]{di2018}%
  \BibitemOpen
  \bibfield  {author} {\bibinfo {author} {\bibfnamefont {M.}~\bibnamefont {Di~Volo}}\ and\ \bibinfo {author} {\bibfnamefont {A.}~\bibnamefont {Torcini}},\ }\href@noop {} {\bibfield  {journal} {\bibinfo  {journal} {Physical review letters}\ }\textbf {\bibinfo {volume} {121}},\ \bibinfo {pages} {128301} (\bibinfo {year} {2018})}\BibitemShut {NoStop}%
\bibitem [{\citenamefont {Capocelli}\ and\ \citenamefont {Ricciardi}(1971)}]{capocelli}%
  \BibitemOpen
  \bibfield  {author} {\bibinfo {author} {\bibfnamefont {R.}~\bibnamefont {Capocelli}}\ and\ \bibinfo {author} {\bibfnamefont {L.}~\bibnamefont {Ricciardi}},\ }\href@noop {} {\bibfield  {journal} {\bibinfo  {journal} {Kybernetik}\ }\textbf {\bibinfo {volume} {8}},\ \bibinfo {pages} {214} (\bibinfo {year} {1971})}\BibitemShut {NoStop}%
\bibitem [{sig()}]{sigmoid_fit}%
  \BibitemOpen
  \href@noop {} {}\bibinfo {note} {We used the function $S(x) = 25 / (1+e^{-k(x-x_0)})$ to approximate the data. The two free parameters, $x_0$ and $k$, were determined by least-squares fitting. We obtained: $x_0=0.32$ and $k=5.67$ for $q=0$, $x_0=0.29$ and $k=5.40$ for $q=0.25$, $x_0=0.22$ and $k=4.94$ for $q=0.5$.}\BibitemShut {Stop}%
\bibitem [{dim()}]{dim}%
  \BibitemOpen
  \href@noop {} {}\bibinfo {note} {Clearly, since ${\rm STA}(\tau)$ is a rate, the integral in time $R$ is dimensionless.}\BibitemShut {Stop}%
\bibitem [{\citenamefont {Gardiner}\ \emph {et~al.}(1985)\citenamefont {Gardiner} \emph {et~al.}}]{gardiner1985handbook}%
  \BibitemOpen
  \bibfield  {author} {\bibinfo {author} {\bibfnamefont {C.~W.}\ \bibnamefont {Gardiner}} \emph {et~al.},\ }\href@noop {} {\emph {\bibinfo {title} {Handbook of stochastic methods}}},\ Vol.~\bibinfo {volume} {3}\ (\bibinfo  {publisher} {springer Berlin},\ \bibinfo {year} {1985})\BibitemShut {NoStop}%
\bibitem [{\citenamefont {Shoham}\ \emph {et~al.}(2006)\citenamefont {Shoham}, \citenamefont {O’Connor},\ and\ \citenamefont {Segev}}]{shoham2006silent}%
  \BibitemOpen
  \bibfield  {author} {\bibinfo {author} {\bibfnamefont {S.}~\bibnamefont {Shoham}}, \bibinfo {author} {\bibfnamefont {D.~H.}\ \bibnamefont {O’Connor}}, \ and\ \bibinfo {author} {\bibfnamefont {R.}~\bibnamefont {Segev}},\ }\href@noop {} {\bibfield  {journal} {\bibinfo  {journal} {Journal of Comparative Physiology A}\ }\textbf {\bibinfo {volume} {192}},\ \bibinfo {pages} {777} (\bibinfo {year} {2006})}\BibitemShut {NoStop}%
\bibitem [{\citenamefont {Hopfield}(1982)}]{hopfield1982neural}%
  \BibitemOpen
  \bibfield  {author} {\bibinfo {author} {\bibfnamefont {J.~J.}\ \bibnamefont {Hopfield}},\ }\href@noop {} {\bibfield  {journal} {\bibinfo  {journal} {Proceedings of the national academy of sciences}\ }\textbf {\bibinfo {volume} {79}},\ \bibinfo {pages} {2554} (\bibinfo {year} {1982})}\BibitemShut {NoStop}%
\bibitem [{\citenamefont {Curti}\ \emph {et~al.}(2004)\citenamefont {Curti}, \citenamefont {Mongillo}, \citenamefont {La~Camera},\ and\ \citenamefont {Amit}}]{curti2004mean}%
  \BibitemOpen
  \bibfield  {author} {\bibinfo {author} {\bibfnamefont {E.}~\bibnamefont {Curti}}, \bibinfo {author} {\bibfnamefont {G.}~\bibnamefont {Mongillo}}, \bibinfo {author} {\bibfnamefont {G.}~\bibnamefont {La~Camera}}, \ and\ \bibinfo {author} {\bibfnamefont {D.~J.}\ \bibnamefont {Amit}},\ }\href@noop {} {\bibfield  {journal} {\bibinfo  {journal} {Neural computation}\ }\textbf {\bibinfo {volume} {16}},\ \bibinfo {pages} {2597} (\bibinfo {year} {2004})}\BibitemShut {NoStop}%
\bibitem [{\citenamefont {Gerstner}\ \emph {et~al.}(2014)\citenamefont {Gerstner}, \citenamefont {Kistler}, \citenamefont {Naud},\ and\ \citenamefont {Paninski}}]{gerstner2014}%
  \BibitemOpen
  \bibfield  {author} {\bibinfo {author} {\bibfnamefont {W.}~\bibnamefont {Gerstner}}, \bibinfo {author} {\bibfnamefont {W.~M.}\ \bibnamefont {Kistler}}, \bibinfo {author} {\bibfnamefont {R.}~\bibnamefont {Naud}}, \ and\ \bibinfo {author} {\bibfnamefont {L.}~\bibnamefont {Paninski}},\ }\href@noop {} {\emph {\bibinfo {title} {Neuronal dynamics: From single neurons to networks and models of cognition}}}\ (\bibinfo  {publisher} {Cambridge University Press},\ \bibinfo {year} {2014})\BibitemShut {NoStop}%
\bibitem [{\citenamefont {Gutkin}(2022)}]{gutkin2022}%
  \BibitemOpen
  \bibfield  {author} {\bibinfo {author} {\bibfnamefont {B.}~\bibnamefont {Gutkin}},\ }in\ \href@noop {} {\emph {\bibinfo {booktitle} {Encyclopedia of computational neuroscience}}}\ (\bibinfo  {publisher} {Springer},\ \bibinfo {year} {2022})\ pp.\ \bibinfo {pages} {3412--3419}\BibitemShut {NoStop}%
\bibitem [{\citenamefont {Zillmer}\ \emph {et~al.}(2007)\citenamefont {Zillmer}, \citenamefont {Livi}, \citenamefont {Politi},\ and\ \citenamefont {Torcini}}]{zillmer2007}%
  \BibitemOpen
  \bibfield  {author} {\bibinfo {author} {\bibfnamefont {R.}~\bibnamefont {Zillmer}}, \bibinfo {author} {\bibfnamefont {R.}~\bibnamefont {Livi}}, \bibinfo {author} {\bibfnamefont {A.}~\bibnamefont {Politi}}, \ and\ \bibinfo {author} {\bibfnamefont {A.}~\bibnamefont {Torcini}},\ }\href@noop {} {\bibfield  {journal} {\bibinfo  {journal} {Physical Review E—Statistical, Nonlinear, and Soft Matter Physics}\ }\textbf {\bibinfo {volume} {76}},\ \bibinfo {pages} {046102} (\bibinfo {year} {2007})}\BibitemShut {NoStop}%
\end{thebibliography}%

\end{document}